\documentclass[table,dvipsnames]{article}
\usepackage{fullpage}
\usepackage{color}
\usepackage[table]{xcolor}
\usepackage{graphicx}
\usepackage{amssymb}
\usepackage{amsmath}
\usepackage{graphicx}
\usepackage{latexsym}
\usepackage{epstopdf}
\usepackage{multicol}
\usepackage{caption}
\usepackage{wrapfig}
\usepackage{pdfpages}
\usepackage[vlined,ruled,commentsnumbered]{algorithm2e}
\usepackage{tikz-qtree,tikz-qtree-compat}
\tikzset{every tree node/.style={align=center, anchor=north}}
\usepackage{array}
\newcolumntype{P}[1]{>{\centering\arraybackslash}p{#1}}
\newcolumntype{M}[1]{>{\centering\arraybackslash}m{#1}}%
\newcolumntype{C}[1]{>{\centering\let\newline\\\arraybackslash\hspace{0pt}}m{#1}}

\usepackage{subcaption}
\usepackage{url}

\usepackage{array}
\let\chapter\undefined

\usepackage{changepage}

\usepackage{listings}
\newcommand*{\defeq}{\mathrel{\vcenter{\baselineskip0.5ex \lineskiplimit0pt
            \hbox{\scriptsize.}\hbox{\scriptsize.}}}%
    =}

\newcolumntype{M}[1]{>{\centering\arraybackslash}m{#1}}

\newcommand{\systemName}{{\tt NLProv}}
\newcommand{\nalir}{{\tt NaLIR}}
\newcommand{\precise}{{\tt PRECISE}}
\newcommand{\selp}{{\tt SelP}}

\newcommand{\trans}{{dependency-to-query-mapping}}
\newcommand{\transu}{{dependency-to-UCQ-mapping}}

\newtheorem{theorem}{Theorem}[section]

\newtheorem{example}[theorem]{Example}

\newtheorem{definition}[theorem]{Definition}

\begin{document}

\title{Explaining Natural Language Query Results}

\author{
  Daniel Deutch\\
  Tel Aviv University\\
  \small{danielde@post.tau.ac.il}
  \and
  Nave Frost\\
  Tel Aviv University\\
  \small{navefrost@mail.tau.ac.il}
  \and
  Amir Gilad\\
  Tel Aviv University\\
  \small{amirgilad@mail.tau.ac.il}
}

\date{}

\maketitle

\begin{abstract}
Multiple lines of research have developed
Natural Language (NL) interfaces for formulating database queries. We build upon this work, but focus on presenting a
highly detailed form of the {\em answers} in NL. The
answers that we present are importantly based on the {\em
provenance} of tuples in the query result, detailing not only the results but also their {\em explanations}. We develop a
novel method for transforming provenance information to NL, by
leveraging the original NL query structure. Furthermore, since
provenance information is typically large and complex, we present
two solutions for its effective presentation as NL text: one that is
based on provenance factorization, with novel desiderata relevant to the NL case, and one that is based on summarization. We have implemented our solution in an end-to-end system supporting questions, answers and provenance, all expressed in NL. Our experiments, including a user study, indicate the quality of our solution and its scalability.
\end{abstract}

\section{Introduction}\label{sec:intro}

In the context of databases, {\bf data provenance}
captures the way in which data is used, combined and manipulated by
the system. Provenance information can for instance be used to
reveal whether data was illegitimately used, to reason about
hypothetical data modifications, to assess the trustworthiness of a
computation result, or to explain the rationale underlying the
computation.

As database interfaces constantly grow in use, in complexity and in the size of data they
manipulate, provenance tracking becomes of paramount
importance. 
In its absence, it is next to impossible to understand the system's operation and to follow the flow of data through the system, which in turn may be extremely harmful for the quality of results.

A setting where the lack of provenance -- and consequently lack of explanations -- is of particular concern, is that of database interfaces geared to be used by non-experts. 
Such non-expert users lack understanding of the system inner workings, and are unable to verify that it has operated correctly. Indeed, an important component of such systems is the interface through which the non-expert communicates her needs/query to the system. But then, how does the system communicate its results back to the non-expert user? And how does it justify it in a manner that the non-expert can understand? 
For each system, developers currently need to develop dedicated solutions, if at all, and we are lacking a generic framework for explanations in this setting.

A particularly flourishing line of work for allowing non-experts to interact with a database, is that of Natural Language Interfaces to Databases (NLIDBs). Multiple such interfaces have been developed in recent years (see {\textit e.g.}
\cite{nalir,Amsterdamer:2015,Kupper1993,SongSSBZBMDDMH15}). The accuracy of translation is constantly improving. Still, it is
far from perfect -- in general, automated translation of free text to a formal language
is an extremely difficult task. Since the users of such systems are typically non-experts, they may have a hard time understanding the result or verifying its correctness. 
Consider for example a complex NL query over a
publication database, of the form ``return all organizations of
authors who published in database conferences after 2005". After
translating this query to SQL and running it using a query engine,
the answer is a list of qualifying organizations. By
looking at the answer, the user has no way of knowing whether the retrieved organizations
really satisfy her specified constraints; a slight error in the
translation process, e.g. misunderstanding ``database conferences"
or erroneously associating ``after 2005" with the conference
inauguration date, could result in a list of organizations
that is completely wrong for the question asked.

In this work we complement the efforts of developing high-quality NLIDBs, by 
developing a generic framework that {\em explains} the results of queries posed to NLIDBs. Explanations are based on provenance, but current provenance models are far too complex to allow for their direct presentation to non-experts. The novelty of our work is that we ``translate" provenance into {\em NL
explanations to the query answers}. The explanations that we provide
elaborate upon answers with additional important information, and
are helpful for understanding {\em why} does each answer qualify to %
the query criteria. 

As an example, consider the Microsoft Academic Search database \cite{mas} and consider the NL
query in Figure \ref{fig:nlcq}. A state-of-the-art NL query engine,
\nalir\ \cite{nalir}, is able to transform this NL query into the
SQL query also shown (as a Conjunctive Query, which is the 
fragment that we focus on in this paper) in Figure \ref{fig:cq}.
When evaluated using a standard database engine, the query returns
the expected list of organizations. However, the answers
(organizations) in the query result lack {\em justification}, which
in this case would include the authors affiliated with each
organization and details of the papers they have published (their
titles, their publication venues and publication years). Such
additional information, corresponding to the notion of {\em
provenance}
\cite{Herschel,why,CheneyProvenance,GKT-PODS07,Greenicdt09} can 
lead to a richer answer than simply providing the
names of organizations: it allows users to also see relevant details
of the qualifying organizations. Provenance information is also
valuable for validation of answers: a user who sees an organization
name as an answer is likely to have a harder time validating that
this organization qualifies as an answer, than if she was presented
with the full details of publications. An understanding of the results also allows users to conclude whether their query was translated correctly and reproduce the results if needed. There are several models of provenance previously suggested in the literature. The {\em tuple-based model} \cite{GKT-PODS07,Greenicdt09} tracks the source tuples which participated in the computation of the results, while the {\em value-based model} \cite{valueProv,DBNotes} is a more fine grained model and follows the values of these tuples.

We propose a novel approach of presenting {\em provenance information for answers of NL queries, again as sentences in Natural Language}. There are several aspects to account for towards a solution, as follows.

\begin{itemize}
\item The provenance model needs to be very detailed. For example, the NL explanations that we aim
for require storing not only which input tuples have contributed to the answer -- in the above example these
may e.g. be the author, organization and publication entries -- but also the way in which they contributed to
the answer. In our example, for generating the required explanations we need to store that the organization
entry has matched the query ``head" and was returned, that the author entry has been joined with it to find
authors of the specific organization, that the publication entry was joined with the author entry, etc. Naturally,
once the query is compiled from NL/examples to e.g. SQL, one could in principle track provenance
as if the query was described in SQL to begin with. As we next explain, this would be sub-optimal.

\item As we shall demonstrate, the way in which the user has phrased the question has a significant impact on
both which parts of the computation needs to be tracked and on the way in which users expect provenance
information to be presented to them. In general, in works on provenance, there is a typically a tight
coupling between the query model and the provenance model. In particular, as we already observed, a suitable way for presenting the
explanations is again as NL sentences, so that we obtain an end-to-end system where questions, answers
and explanations are all expressed in Natural Language. Thus, the provenance model needs to
keep track of which parts of the NL question have contributed to which parts of the computation. 
Furthermore, We use the value-based model of provenance as it is the implicitly enforced by the NLIDB which maps words to variables. Once we have this mapping, we store the mappings between variables to values to be able to reverse it.
A major challenge in this respect is to design the model so that it correctly captures those parts of the provenance
that are ``important'' based on the user question.
As the basis for our provenance model, we use the value-based model of provenance (as opposed to tuple-based) as it is the implicitly enforced by the NLIDB which maps words to variables in the query. Once we have this mapping, we store the mappings between variables to values to be able to assemble an explanation sentence.

\item Then, given the tracked provenance, we further need to translate it back from the formal model to an NL sentence. Generating NL sentences is a difficult task in general -- but importantly, here we have the NL question that can guide us. A challenge is then how to ``plug-in" different parts of the provenance back into the NL question, to obtain a coherent, well-formed answer.  
\item Last, we need to address the challenge of provenance size. In particular, full information about the manner in which a result is obtained from the input
data (and even full description of the input data itself) is typically exhaustively long to present, especially to a non-expert.      
\end{itemize}

The end result for our running example is demonstrated in  Figure \ref{fig:single}, which shows one of the explained answers outputted by our system in response to the NL query in Figure \ref{fig:nlcq}.
    
Having explained the overall approach and challenges, we next provide more details on each of our key contributions.

\begin{figure}
    \begin{center}
    \begin{subfigure}[b]{1.0\columnwidth}
        \centering
        \small{
        \begin{tabular}{|l|}
            \hline
            \verb"return the organization of authors who published papers"\\
            \verb"in database conferences after 2005"\\
            \hline
        \end{tabular}
        \caption{NL Query}\label{fig:nlcq}
    }
    \end{subfigure}\hfill%

    \begin{subfigure}[b]{1.0\columnwidth}
        \centering
        \small{
        \begin{tabular}{|l|}
            \hline
            \verb"query(oname) :- org(oid, oname), conf(cid, cname), "\\
            \verb"pub(wid, cid, ptitle, pyear), author(aid, aname, oid), "\\
            \verb"domainConf(cid, did), domain(did, dname), "\\
            \verb"writes(aid, wid), dname = 'Databases', pyear > 2005"\\
            \hline
        \end{tabular}
        \caption{CQ $Q$}\label{fig:cq}
    }
    \end{subfigure}\hfill%
    \caption{NL Query and CQ $Q$} \label{fig:nlquery}
    \end{center}
\end{figure}

\begin{figure}
    \begin{center}
        \small{
            \begin{tabular}{|l|}
                \hline
                \verb"TAU is the organization of Tova M. who published 'OASSIS...' in SIGMOD in 2014"\\
                \hline
            \end{tabular}
        }
    \caption{Answer for a Single Assignment} \label{fig:single}
    \end{center}
\end{figure}

\paragraph{Provenance Tracking Based on the NL Query Structure}  As mentioned above, a first key idea in our solution is to leverage the {\em NL query
    structure} in constructing NL provenance. Our solution is generic in that it is not specific to a concrete NL interface (we do have some requirements on the operation of the underlying interface, as we detail below). In our implementation, we use and modify \nalir\footnote{We are
    extremely grateful to Fei Li and H.V. Jagadish for generously
    sharing with us the source code of \nalir, and providing invaluable
    support.} so that we store exactly which parts of the NL query
translate to which parts of the formal query. Then, we evaluate the
formal query using a provenance-aware engine (we use \selp\ \cite{vldb15deutch}),
further modified so that it stores which parts
of the query ``contribute" to which parts of the provenance. By
composing these two ``mappings" (text-to-query-parts and
query-parts-to-provenance) we infer which parts of the NL query text
are related to which provenance parts. Finally, we use the
latter information in an ``inverse" manner, to translate the
provenance to NL text.

\paragraph{Factorization} A second key idea is related to the provenance size. In typical
scenarios, a single answer may have multiple explanations (multiple
authors, papers, venues and years in our example). A na{\"i}ve solution
is to formulate and present a separate sentence corresponding to
each explanation. The result will however be, in many cases, very
long and repetitive. As observed already in previous work
\cite{Chapman2008,Olteanu2012}, different assignments (explanations)
may have significant parts in common, and this can be leveraged in a
{\em factorization} that groups together multiple occurrences. In
our example, we can {\textit e.g.} factorize explanations based on author, 
paper name, conference name
or year. Importantly, we impose a novel constraint on the factorizations that we look for (which we call {\em compatibility}), intuitively capturing that their structure is consistent with
a partial order defined by the parse tree of the question. This constraint is needed so that we can translate the factorization back to an NL answer whose structure is similar to that of the question. Even with this constraint, there may still be exponentially many (in the size of the provenance expression) compatible factorizations, and we look for the factorization with minimal size out of the compatible ones; for comparison, previous work looks for the minimal factorization with no such ``compatibility constraint". The corresponding decision problem remains coNP-hard (again in the provenance size), but we devise an effective and simple greedy solution.  We further
translate factorized representations to concise NL sentences, again leveraging the structure of the NL query.

\paragraph{Summarization} We propose {\em summarized} explanations by replacing details of different parts of the explanation by
their synopsis, {\textit e.g.} presenting only the number of papers published by each author, the number of authors, or the overall number of papers published by authors of each organization.
Such summarizations incur by nature a loss of information but are
typically much more concise and easier for users to follow. Here
again, while provenance summarization has been studied before ({\textit e.g.} \cite{ainy,Re2008}), the desiderata of a summarization needed for NL sentence generation are different, rendering previous
solutions inapplicable here. We observe a tight correspondence between factorization and summarization: every factorization gives rise to multiple possible summarizations, each
obtained by counting the number of sub-explanations that are ``factorized together". 
We provide a robust solution,
allowing to compute NL summarizations of the provenance, of varying
levels of granularity.

\paragraph{Implementation and Experiments} We have implemented our solution in a system prototype called \systemName\ \cite{nlprov},
forming an end-to-end NL interface to database querying where the NL queries, answers and provenance information are all expressed in NL.
We have further conducted extensive experiments whose results indicate the scalability of the solution as well as the quality of the results, the latter through a user study. 

This paper is an extended version of our PVLDB 2017 paper \cite{DeutchFG17} and includes a new section on the translation of provenance to NL for UCQs, a new section on a generalized solution that is not specific to \nalir\ and a discussion of the use of other provenance models, new and comprehensive experiments, and an extended in-depth review of related work.

\section{Preliminaries}\label{sec:prelim}

We provide here the necessary preliminaries on Natural Language Processing, conjunctive queries and provenance.

\subsection{NL and Formal Queries}

We start by recalling some basic notions from NLP, as they pertain
to the translation process of NL queries to a formal query language. We further recall a particular formal query language of interest, namely Union of Conjunctive Queries.

A key notion that we will use is that of the {\em syntactic
dependency tree} of NL queries:

\begin{definition}
    A dependency tree $T=(V,E,L)$ is a node-labeled tree where labels consist of two components, as follows: (1) Part of Speech ($POS$): the syntactic role of the word \cite{klein2003,Marcus1993} ; (2) Relationship ($REL$): the grammatical relationship between the
    word and its parent in the dependency tree \cite{lrec2006}.
\end{definition}

\begin{figure}
    \hspace*{-0.6cm}
    \centering
    \begin{subfigure}[b]{0.5\columnwidth}
        \centering
        \begin{tikzpicture}[scale=0.9, level 1/.style={level distance=0.7cm}, level 3/.style={sibling distance=-3mm, level distance=0.7cm},
        level 2/.style={level distance=0.7cm},level 4/.style={level distance=0.1cm},
        snode/.style = {shape=rectangle, rounded corners, draw, align=center, top color=white, bottom color=blue!20}]
        \Tree 
        [.\node[snode] {\small{return}};
        [.\node[snode] {\small{object}};
        [\qroof{\small{others}}. ]
        [.\node[snode] {\small{verb mod}}; \node[snode] {\small{nsubj}};
        [\qroof{\small{properties}}. ] ]
        ]
        ]
        \end{tikzpicture}
        \caption{Verb Mod.}\label{fig:abstParse1}
    \end{subfigure}\hfill%
    \begin{subfigure}[b]{0.4\columnwidth}
        \centering
        \begin{tikzpicture}[scale=0.9, level 1/.style={level distance=0.7cm}, level 2/.style={level distance=0.7cm},
        level 3/.style={level distance=0.7cm},,level 4/.style={level distance=0.1cm},
        snode/.style = {shape=rectangle, rounded corners, draw, align=center, top color=white, bottom color=blue!20}]
        \Tree 
        [.\node[snode] {\small{return}};
        [.\node[snode] {\small{object}};
        [\qroof{\small{others}}. ]
        [.\node[snode] {\small{non-verb mod}};
        [\qroof{\small{properties}}. ]
        ]
        ]
        ]
        \end{tikzpicture}
        \caption{Non-Verb Mod.}\label{fig:abstParse2}
    \end{subfigure}
    \caption{Abstract Dependency Trees}\label{fig:abs}
\end{figure}
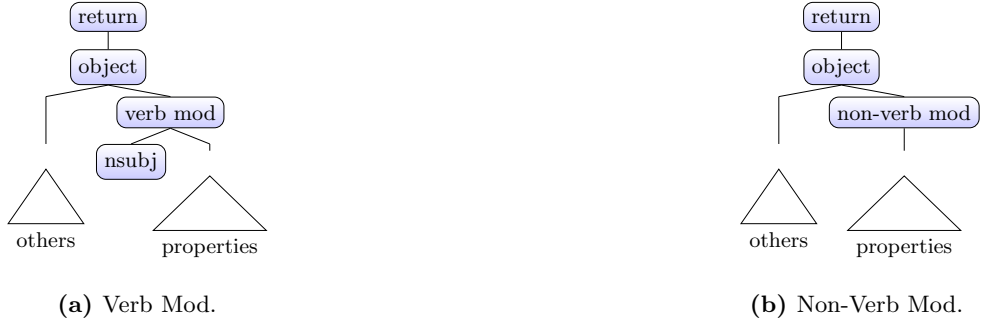

We focus on a sub-class of queries handled by \nalir, namely that of Union of Conjunctive Queries, possibly with comparison operators ($=,>,<$) and logical combinations thereof (\nalir\ further supports nested queries and aggregation). Formally, fix a database schema, i.e. a set of relation names along with their arities (number of attributes). A query is then defined with respect to a schema.

\begin{definition} (From \cite{abiteboul1995foundations})
    A Union of Conjunctive Queries $Q$ is a set of Conjunctive Queries $Q_i$. 
    
    In turn, a conjunctive query is an expression $ans(u) \longleftarrow R1(u1),...,Rn(un),C$
where $R1,...,Rn$ are relation names in the database schema, and
$u,u1,...,un$ are tuples with either variables or constants, with $ui$ conforming to the schema of $Ri$. Variables in $u$ must appear in at least one of $u1,...un$. Finally, $C$ is a sequence of comparison constraints ($=,>,<$) over variables in $u1,...un$ and constants.   
\end{definition}

The corresponding NL queries in \nalir\ follow one of the two (very general) abstract forms described in Figure \ref{fig:abs}: 
an object (noun) is sought for, that satisfies some properties, possibly described through a complex sub-sentence rooted by a {\em modifier} (which may or may not be a verb, a distinction whose importance will be made clear in our algorithms that follow).

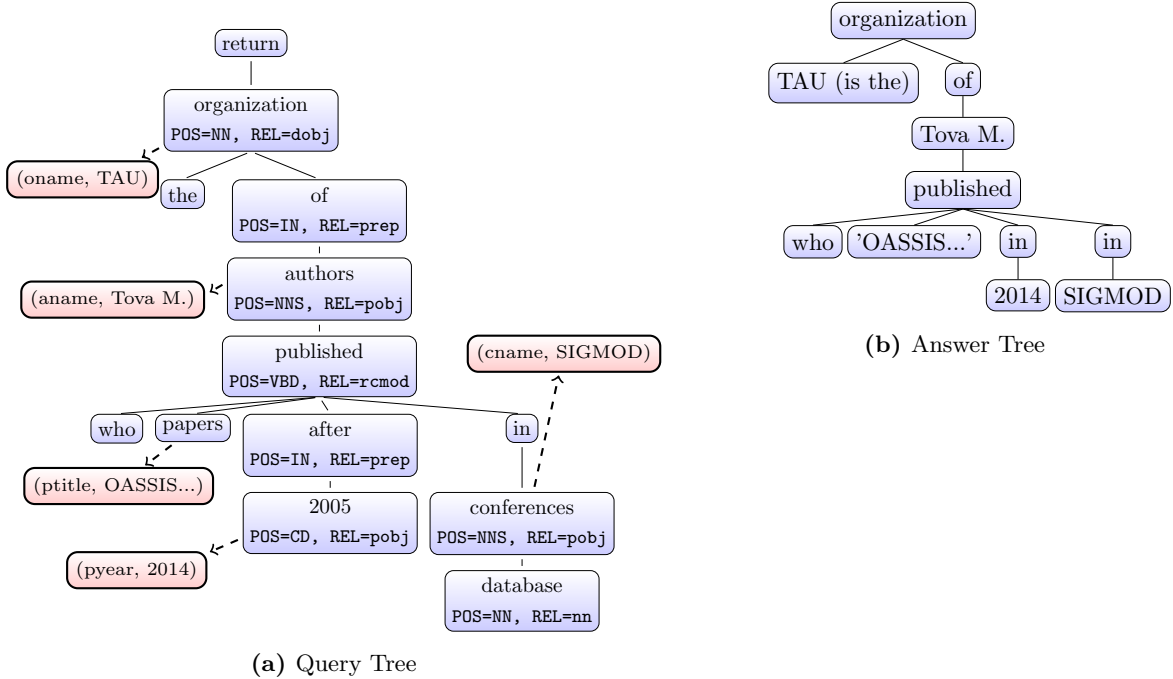
\begin{figure}[htb!]
    \begin{subfigure}{.7\textwidth}
        \centering
        \begin{tikzpicture}[thick, scale=.8,
        ssnode/.style={fill=mygreen,node distance=.1cm,draw,circle, minimum size=.1mm},
        -,shorten >= 2pt,shorten <= 2pt,
        level 1/.style={level distance=0.2cm},
        every node/.style = {shape=rectangle, rounded corners, draw, align=center, top color=white, bottom color=blue!20},
        map/.style={shape=rectangle, rounded corners, draw, align=center, top color=white, bottom color=red!20},
        level 2/.style={level distance=1.5cm, sibling distance=-19mm},
        level 5/.style={sibling distance=2mm},
        level distance=1.3cm
        ]
        \node[map] (b5) at (-4.5,4.5) {\scriptsize{(oname, TAU)}};
        \node[map] (b1) at (-4,2.5) {\scriptsize{(aname, Tova M.)}};
        \node[map] (b2) at (-3.9,-0.6) {\scriptsize{(ptitle, OASSIS...)}};
        \node[map] (b3) at (3.5,1.6) {\scriptsize{(cname, SIGMOD)}};
        \node[map] (b4) at (-3.6,-2) {\scriptsize{(pyear, 2014)}};

        \begin{scope}[level 1/.style={level distance=1.0cm}, yshift= 7cm, xshift= -1.7cm]

        \Tree 
        [.return
        [.\node[] (a5) {organization\\ \small {\tt POS=NN, REL=dobj}}; {the}
        [.\node[] (z) {of \\ \small {\tt POS=IN, REL=prep} };
        [.\node[] (a1) {authors\\ \small {\tt POS=NNS, REL=pobj}};
        [.{published \\ \small {\tt POS=VBD, REL=rcmod} } {who}
        [.\node[] (a2) {papers}; ]
        [.{after \\ \small {\tt POS=IN, REL=prep}} [.\node[] (a4) {2005 \\ \small {\tt POS=CD, REL=pobj}}; ] ]
        [.{in}
        [.\node[] (a3) {conferences \\ \small {\tt POS=NNS, REL=pobj}};
        [.\node[] (a6) {database \\ \small {\tt POS=NN, REL=nn} }; ]
        ]
        ]
        ]
        ]
        ]
        ]
        ]

        \end{scope}

        \draw[->,thick,shorten <=2pt,shorten >=2pt, dashed] (a5) -- (b5);
        \draw[->,thick,shorten <=2pt,shorten >=2pt, dashed] (a1) -- (b1);
        \draw[->,thick,shorten <=2pt,shorten >=1, dashed] (a2) -- (b2);
        \draw[->,thick,shorten <=2pt,shorten >=1, dashed] (a3) -- (b3);
        \draw[->,thick,shorten <=2pt,shorten >=1, dashed] (a4) -- (b4);
        \end{tikzpicture}
        \caption{Query Tree}\label{fig:parseMap}
    \end{subfigure}%
    \begin{subfigure}[b]{.3\textwidth}
        \centering
        \begin{tikzpicture}[scale=.9,level 1/.style={level distance=0.9cm, sibling distance=-20mm}, level 2/.style={sibling distance=-20mm}, level distance=.8cm,
        every node/.style = {shape=rectangle, rounded corners, draw, align=center, top color=white, bottom color=blue!20}]
        
        
        \Tree
        [.organization {TAU (is the)}
        [.of
        [.{Tova M.}
        [.published who {'OASSIS...'}
        [.in 2014 ]
        [.in SIGMOD ]
        ]
        ]
        ]
        ]
        
        \end{tikzpicture}
        \caption{Answer Tree}\label{fig:organswertree}
    \end{subfigure}
    \caption{Question and Answer Trees}\label{fig:trees}
\end{figure}

\begin{example}\label{ex:labels}
    Reconsider the NL query in Figure \ref{fig:nlcq}; its dependency
    tree is depicted in Figure \ref{fig:parseMap} (ignore for now the
    arrows). The part-of-speech (POS) tag of each node reflects its
    syntactic role in the sentence -- for instance, ``organization" is a
    noun (denoted ``NN"), and ``published" is a verb in past tense
    (denoted ``VBD"). The relation (REL) tag of each node reflects the
    semantic relation of its sub-tree with its parent. 
    For instance, the REL of ``of'' is prep (``prepositional modifier") meaning
    that the sub-tree rooted at ``of" describes a property of
    ``organization" while forming a complex sub-sentence.
    The tree in Figure \ref{fig:parseMap} matches the abstract tree in Figure \ref{fig:abstParse2} since
    ``organization" is the object and ``of" is a non-verb modifier (its POS tag is IN, meaning ``preposition or subordinating conjunction")  rooting a sub-sentence describing ``organization".
\end{example}

The dependency tree is transformed by \nalir, based also on schema knowledge, to SQL. We focus in this work on NL queries that are compiled into Union of Conjunctive Queries (UCQs), and discuss extensions to aggregates and nested queries below.  

\begin{example}
    \label{ex:translation}
    Reconsider our running example NL query in Figure \ref{fig:nlcq}; a counterpart Conjunctive Query is shown in Figure \ref{fig:cq}.
    Some words of the NL query have been mapped by \nalir\ to variables in the query, {\textit e.g.}, the word ``organization" corresponds to the head variable ($oname$).
    Additionally, some parts of the sentence have been complied to boolean conditions based on the MAS schema, {\textit e.g.}, the part ``in database conferences" was translated
    to $dname =$ `Databases' in the CQ depicted in Figure \ref{fig:cq}. Figure \ref{fig:parseMap} shows the mapping of some of the nodes in the NL query dependency tree to variables of $Q$ (ignore for now the values next to these variables).
\end{example}

The translation performed by \nalir\ from an NL query to a formal one can be captured by a {\em mapping} from (some) parts of the sentence to parts of the formal query. This mapping is not a novel contribution of this paper, but we will employ this mapping to generate the NL explanation.

\begin{definition}\label{def:translation}
    Given a dependency tree $T=(V,E,L)$ and a CQ $Q$, a \trans\\$\tau: V \to Vars(Q)$ is a partial function mapping a subset of the
    dependency tree nodes to the variables of $Q$.
\end{definition}

\subsection{Provenance}

After compiling a formal query corresponding to the user's NL query,
we evaluate it and keep track of {\em provenance}, to be used in
explanations. 

As explained in Section \ref{sec:intro}, there are essentially two explanation models that will come into play here. The first is a provenance model for the underlying formal query, in our case Union of Conjunctive Queries. We next discuss existing provenance models, then choose a particular model that fits our construction.  The second, which we will discuss below, is coupled with the Natural Language model.

In terms of provenance for formal database queries, previous work has proposed a large number of different models (see Section 7 for an overview of related work). A basic distinction that we already need to make is between fine-grained and coarse-grained provenance. Generally speaking, the former keeps track of which tuples (or even cells) have contributed to each result, while the latter keeps track of the general input and output of each query/workflow operator, without necessarily connecting each input and output pieces. Here our goal is to explain individual query results, and so a fine-grained provenance model is sought for. 

In context of database queries, capturing fine-grained provenance means that we keep track of the {\em assignments} of database tuples to query atoms. Assignments are the basic building block of query evaluation, and for UCQs they are defined as follows:     
\begin{definition} \label{def:assignment}
    An assignment $\alpha$ for a query $Q \in CQ$ with respect to a
    database instance $D$ is a mapping of the relational atoms of $Q$ to
    tuples in $D$ that respects relation names and induces a mapping
    over variables/constants, {\textit i.e.} if a relational atom $R(x_1, ..., x_n)$ is mapped to a tuple $R(a_1, ..., a_n)$ then we say that
    $x_i$ is mapped to $a_i$ (denoted $\alpha(x_i) = a_i$, overloading notations) and we
    further say that the tuple was {\em used} in $\alpha$. We require
    that any variable will be mapped to a single value,
    and that any constant will be mapped to itself. We further define
    $\alpha(head(Q))$ as the tuple obtained from $head(Q)$ by replacing
    each occurrence of a head variable $x_i$ by $\alpha(x_i)$. The set of assignments to $Q$ with respect to $D$ is denoted by $\Gamma(Q,D)$. Note that a single tuple in the query result may have been obtained by multiple assignments in $\Gamma(Q,D)$.
    
    Then, for a UCQ $Q$ whose CQs are $Q_1,...,Q_n$, the set of assignments to $Q$ is defined as the union of the sets of assignments to its CQs, namely $\Gamma(Q)=\bigcup\limits_{i=1}^n\Gamma(Q_i)$. The notion of a tuple being obtained from an assignment and the $\alpha$ notation immediately extend, noting that a single tuple may be obtained from assignments to multiple conjunctive queries.
\end{definition}

Assignments allow for defining the semantics of CQs: a tuple $t$
is said to appear in the query output if there exists an assignment
$\alpha$ s.t. $t=\alpha(head(Q))$. They will also be useful in
defining provenance below.

\begin{example}\label{ex:cqmap}
    Consider again the query $Q$ in Figure
    \ref{fig:cq} and the database in Figure \ref{tbl:dblp}. 
    The tuple (TAU) is an output of $Q$ when assigning the highlighted tuples to the atoms of $Q$.
    As part of this assignment, the tuple (2, TAU) (the second tuple in the $org$ table) and (4, Tova M., 2) (the second tuple of the $author$
    table) are assigned to the first and second atom of $Q$, respectively.
    In addition to this assignment, there are 4 more assignments that produce the tuple (TAU) and one assignment that produces the tuple (UPENN).
\end{example}

\begin{figure}
    \begin{center}
    \small{
        \begin{tabular}{|l|}
            \hline
            \verb"(oname,TAU)"$\cdot$\verb"(aname,Tova M.)"$\cdot$\verb"(ptitle,OASSIS...)"$\cdot$\verb"(cname,SIGMOD)"$\cdot$\verb"(pyear,14')"$+$ \\
            \verb"(oname,TAU)"$\cdot$\verb"(aname,Tova M.)"$\cdot$\verb"(ptitle,Querying...)"$\cdot$\verb"(cname,VLDB)"$\cdot$\verb"(pyear,06')"$+$ \\
            \verb"(oname,TAU)"$\cdot$\verb"(aname,Tova M.)"$\cdot$ \verb"(ptitle,Monitoring..)"$\cdot$\verb"(cname,VLDB)"$\cdot$\verb"(pyear,07')"$+$\\
            \verb"(oname,TAU)"$\cdot$\verb"(aname,Slava N.)"$\cdot$\verb"(ptitle,OASSIS...)"$\cdot$\verb"(cname, SIGMOD)"$\cdot$\verb"(pyear,14')"$+$ \\
            \verb"(oname,TAU)"$\cdot$\verb"(aname,Tova M.)"$\cdot$\verb"(ptitle,A sample...)"$\cdot$\verb"(cname,SIGMOD)"$\cdot$\verb"(pyear,14')"$+$\\
            \verb"(oname,UPENN)"$\cdot$\verb"(aname,Susan D.)"$\cdot$\verb"(ptitle,OASSIS...)"$\cdot$\verb"(cname,SIGMOD)"$\cdot$\verb"(pyear,14')"\\
            \hline
        \end{tabular}
    }
    \caption{Value-level Provenance} \label{fig:orgprov}
    \end{center}
\end{figure}

Assignments may be used in provenance in multiple ways, varying in their granularities. For instance, the lineage \cite{Trio} of a result tuple $t$ is the set of input tuples appearing in some assignment yielding $t$; the why-provenance of $t$ is the set of sets of tuples participating in such assignments, i.e. the contributing tuples are grouped        
based on the assignments they are used in. 
These approaches were shown in \cite{GKT-PODS07} to be concrete examples of a general algebraic construction, termed semiring provenance. At a high-level, the idea there is that we introduce two symbolic operations $``+"$ and $``\cdot"$, and use them to form algebraic representations of the provenance. Concretely, $``+"$ is used for alternative derivations and $``\cdot"$ is used for combined derivation: for a given output tuple, we sum over the assignments that have yielded it, and each assignment is represented via a  multiplication over the terms that has contributed to it.  The idea is that assignments capture the
{\em reasons} for a tuple to appear in the query result, with each
assignment serving as an {\em alternative} such reason (indeed, the
existence of a single assignment yielding the tuple suffices,
according to the semantics, for its inclusion in the query result).

In \cite{GKT-PODS07}, the basic atomic units that appear in a provenance expression are the ``annotations" (intuitively identifiers, or meta-data associated with the tuples) of the {\em tuples} that contribute to an assignment. Here, in order to form a detailed explanation of the result of an NL query, we need to keep track of a finer-grained resolution. Within each assignment, we keep record of the {\em value} assigned to each
variable, and note that the {\em conjunction} of these value
assignments is required for the assignment to hold.  

\begin{definition}\label{def:valuelevel}
    \label{def:prov}
    Let $A(Q,D)$ be the set of assignments for a UCQ Q and a database instance $D$. We define the {\em value-level provenance} of $Q$ w.r.t. $D$ as
    $$\sum_{\alpha \in A(Q,D)}\Pi_{\{x_i,a_i \mid \alpha(x_i) = a_i\}}(x_i,a_i)$$.
\end{definition}

The reason for our use of a value-based rather than tuple-based provenance model is that, as we will next show, we wish to connect different pieces of the provenance back to the NL question, in order to form a detailed NL explanation.

\begin{figure}[!htb]
    \centering \scriptsize
    \begin{minipage}{.5\linewidth}
        \centering
        \caption*{Rel. $\text{\textit{org}}$}\label{tbl:organization}
        \begin{tabular}{| c | c | c | c | c | c |}
            \hline oid & oname \\
            \hline 1 & UPENN \\
            \rowcolor{lightgray}
            \hline 2 & TAU \\
            \hline
        \end{tabular}
        
    \end{minipage}%
    \begin{minipage}{.5\linewidth}
        \centering
        \caption*{Rel. $\text{\textit{author}}$}\label{tbl:author}
        \begin{tabular}{| c | c | c | c | c | c |}
            \hline aid & aname & oid \\
            \hline 3 & Susan D. & 1 \\
            \rowcolor{lightgray}
            \hline 4 & Tova M. & 2 \\
            \hline 5 & Slava N. & 2 \\
            \hline
        \end{tabular}
        
    \end{minipage}

    \begin{minipage}{0.7\linewidth}
        \centering
        \caption*{Rel. $\text{\textit{pub}}$}\label{tbl:publication}
        \begin{tabular}{| c | c | c | c | c | c |}
            \hline wid & cid & ptitle & pyear\\
            \rowcolor{lightgray}
            \hline 6 & 10 & ``OASSIS..." & 2014 \\
            \hline 7 & 10 & ``A sample..." & 2014 \\
            \hline 8 & 11 & ``Monitoring..." & 2007 \\
            \hline 9 & 11 & ``Querying..." & 2006 \\
            \hline
        \end{tabular}
        
    \end{minipage}%
    \begin{minipage}{.3\linewidth}
        \centering
        \caption*{Rel. $\text{\textit{writes}}$}\label{tbl:writes}
        \begin{tabular}{| c | c | c | c | c | c |}
            \hline aid & wid\\
            \rowcolor{lightgray}
            \hline 4 & 6 \\
            \hline 3 & 6 \\
            \hline 5 & 6 \\
            \hline 4 & 7 \\
            \hline 4 & 8 \\
            \hline 4 & 9 \\
            \hline
        \end{tabular}
        
    \end{minipage}

    \begin{minipage}{.33\linewidth}
        \centering
        \caption*{Rel. $\text{\textit{conf}}$}\label{tbl:conference}
        \begin{tabular}{| c | c | c | c | c | c |}
            \hline cid & cname\\
            \rowcolor{lightgray}
            \hline 10 & SIGMOD \\
            \hline 11 & VLDB \\
            \hline
        \end{tabular}
        
    \end{minipage}
    \begin{minipage}{.33\linewidth}
        \centering
        \caption*{Rel. $\text{\textit{domainConf}}$}\label{tbl:domainconference}
        \begin{tabular}{| c | c | c | c | c | c |}
            \hline cid & did\\
            \rowcolor{lightgray}
            \hline 10 & 18 \\
            \hline 11 & 18 \\
            \hline
        \end{tabular}
        
    \end{minipage}%
    \begin{minipage}{.33\linewidth}
        \centering
        \caption*{Rel. $\text{\textit{domain}}$}\label{tbl:domain}
        \begin{tabular}{| c | c | c | c | c | c |}
            \hline did & dname\\
            \rowcolor{lightgray}
            \hline 18 & Databases \\
            \hline
        \end{tabular}
        
    \end{minipage}

    \caption{DB Instance}\label{tbl:dblp}
\end{figure}

\begin{example}\label{ex:provbasic}
    Re-consider our running example query and consider the database in
    Figure \ref{tbl:dblp}. The value-level provenance is shown in
    Figure \ref{fig:orgprov}. Each of the 6
    summands stands for a different assignment ({\textit i.e.} an alternative reason for the tuple to appear in the result). Assignments are represented as multiplication of pairs
    of the form $(var, val)$ so that $var$ is assigned $val$ in the
    particular assignment. We only show here variables to which a query
    word was mapped; these will be the relevant variables for formulating
    the answer.
\end{example}

It is important to note that we refer to provenance as the mapping between the variables in the query to the values in the database which occurs during the evaluation process. This process is completely separate from \nalir's framework. The provenance is stored as part of the evaluation of the formal query inferred by \nalir\ over the database, and is therefore performed after \nalir\ has completed the query inference process.

\section{First step: Conjunctive Queries and a Single Assignment} \label{sec:single}

We now start describing our transformation of provenance to an NL sentence, leveraging the structure of the original question. We focus in this section on the case of a Conjunctive Query and a single assignment to its clauses. In subsequent sections we show how to extend the solution to multiple assignments and unions of conjunctive queries, where the solution presented in this section will serve as a building block. 

\subsection{Mapping NL to Provenance and Back}

Our first important observation is that {\em words} in the NL question can be connected to {\em (variable,value) pairs} in the provenance polynomial. For instance, ``conference" corresponds to the assignment of $cname$ to $SIGMOD$ or $VLDB$, ``author" corresponds to the assignment of $aname$ to $Tova M.$, and so on. The reason this connection is important is that it gives us strong hints on how to form a detailed answer in Natural Language: given this information we know for instance that $SIGMOD$ should replace/reside next to ``database conferences'' (the decision of which of the two options to follow will be discussed below based on the sentence structure).     
Fortunately, the choice of models we have made in the preliminaries gives us relatively straightforward means to derive this mapping. 
The idea is to marry the two mappings discussed in the previous section as a step towards generating an NL explanation: the \trans\ performed by \nalir\ and the value-based provenance to get a direct mapping from words to database values. 
First, we have the \trans\ mapping (Definition \ref{def:translation}) from the NL query's dependency tree (e.g. ``author") to
query variables (e.g. ``aname"), which we get from the NLIDB. Second, we have, in the value-based provenance, a detailed account of the assignments of query variables to values from the database (e.g. ``aname" to ``Tova. M."). If we compose this mapping, we get a (partial) mapping from words in the NL question to data values.

\begin{example}\label{ex:nlmap}
	Continuing our running example, consider the assignment represented by the first monomial of Figure \ref{fig:orgprov}. Further reconsider Figure \ref{fig:parseMap}, and now note that each node is associated with a pair $(var,val)$ of the variable to which the node was mapped, and the value that this variable was assigned in this particular assignment. For instance, the node ``organization" was mapped to the variable $oname$ which was assigned the value ``TAU".
\end{example}

\subsection{Building an Answer Tree}

Having established the mapping from words in the NL query to values in the provenance, we are ready to form a basic tree for the provenance-aware answer. The idea is now to follow the structure of the NL query dependency tree
and generate an answer tree with the same structure by replacing/modifying
the words in the question with the values from the result and provenance
that were mapped using the \trans\ and the assignment. Yet, note that
simply replacing the values does not always result in a
coherent sentence, as shown in the following example.

\begin{example}

    Re-consider the dependency tree depicted in Figure \ref{fig:parseMap}.
    If we were to replace the value in the organization node to the value ``TAU" mapped to it,
    the word ``organization" will not appear in the answer although it is needed to produce the
    coherent answer depicted in Figure \ref{fig:single}.
    Without this word, it is unclear how to deduce the information about the connection
    between ``Tova M." and ``TAU".

\end{example}

We next account for these difficulties and present an algorithm that outputs the dependency tree of a detailed answer, under some plausible assumptions on the structure of the question tree.  

Recall that we have assumed that the dependency tree of the NL query follows one of the abstract forms in Figure \ref{fig:abs}. We distinguish between two cases based on nodes whose $REL$ (relationship with parent node) is {\em modifier}; in the first case, the clause begins with a verb modifier ({\textit e.g.}, the node ``published'' in Fig. \ref{fig:parseMap} is a verb modifier)
and in the second, the clause begins with a non-verb modifier ({\textit e.g.}, the node ``of'' in Fig. \ref{fig:parseMap} is a non-verb modifier). Algorithm \ref{algo:ansTree} considers these two forms of dependency tree and provides a
tailored solution for each one in the form of a dependency tree that fits the correct answer structure.
It does so by augmenting the query dependency tree into an answer tree.

\IncMargin{1em}
\begin{algorithm}[!htb]
    \SetKwFunction{ComputeAnswerTree}{ComputeAnswerTree}
    \SetKwInOut{Input}{input}\SetKwInOut{Output}{output}
    \LinesNumbered
    \Input{A dependency tree $T_Q$, an answer tree $T_A$ (empty in the first call), a \trans\ $\tau$, an assignment $\alpha$, a node $object \in T_Q$}
    \Output{Answer tree with explanations $T_A$} \BlankLine

    $child \defeq null$\;
    \If {$object$ is a leaf}
    {\label{line:leaf}
        $value = \alpha(\tau(object))$\; \label{line:mapU}
        $Replace(object, value, T_A)$\; \label{line:swapUnon}
    }

    \ElseIf {$L(object).REL$ is mod}
    {\label{line:mod}
        $value = \alpha(\tau(child_{T_Q} (object)))$\; \label{line:mapChildU}
        $Replace(tree(object), value, T_A)$\;\label{line:swapU}
        $AddParent(T_A, value)$\;
    }

    \ElseIf {$object$ has a child $v$ s.t. $L(v).REL \in MOD$ and $L(v).POS \notin VERB$}
    {\label{line:noVerbMod}
        $Adjust(T_Q, T_A, \tau, \alpha, object, false)$\; \label{line:initfalse}
        $child \defeq v$\;
    }

    \ElseIf {$object$ has a child $v$ s.t. $L(v).REL \in MOD$ and $L(v).POS \in VERB$}
    {\label{line:verbMod}
        $Adjust(T_Q, T_A, \tau, \alpha, object, true)$\;
        $child \defeq v$\;
    }

    \If {$child \neq null$}
    {\label{line:hasMod}
        \ForEach {$u \in children_{T_Q}(child)$}
        {\label{line:childV}
            $ComputeAnswerTree(T_Q, T_A, \tau, \alpha, u)$\;
        }
    }

    \Return $T_A$\; \label{line:return2}

    \caption{ComputeAnswerTree} \label{algo:ansTree}
\end{algorithm} \DecMargin{1em}

The algorithm operates as follows.
We start with the dependency tree of the NL query, an empty answer tree $T_A$, a \trans\, an assignment and a node $object$ from the query tree.
We denote the set of all modifiers by $MOD$ and the set of all verbs by $VERB$.
The algorithm is recursive and handles several cases, depending on $object$ and its children in the dependency tree.
If the node $object$ is a leaf (Line \ref{line:leaf}), we replace it with the value mapped to it by \trans\ and the assignment, if such a mapping exists. Otherwise (it is a leaf without a mapping), it remains in the tree as it is.
Second, if $L(object).REL$ is a modifier (Line \ref{line:mod}), we call the procedure $Replace$ in order to replace its entire subtree with the value mapped to it and add the suitable word for equality,
depending on the type of its child ({\textit e.g.}, location, year, etc. taken from the pre-prepared table), as its parent (using procedure $AddParent$).
The third case handles a situation where $object$ has a non-verb modifier child (Line \ref{line:noVerbMod}). We use the procedure $Adjust$ with a $false$ flag to copy $T_Q$ into $T_A$,
remove the $return$ node and add the value mapped to $object$ as its child in $T_A$.
The difference in the fourth case (Line \ref{line:verbMod}) is the value of $flag$ is now $true$.
This means that instead of adding the value mapped to $object$ as its child, the $Adjust$ procedure replaces the node with its value.
Finally, if $object$ had a modifier child $child$ (verb or non-verb), the algorithm makes a recursive call for all of the children of $child$ (Line \ref{line:childV}). This recursive call is
needed here since a modifier node can be the root of a  complex sub-tree (recall Example \ref{ex:labels}).

\begin{example}\label{ex:singleDerSentence}

    Re-consider Figure \ref{fig:parseMap}, and note the mappings from the nodes to the variables and values as reflected in the boxes next to the nodes.
    To generate an answer, we follow the NL query structure, ``plugging-in" mapped database
    values. We start with ``organization", which is the first $object$ node.
   Observe that ``organization" has the child ``of" which is
    a non-verb modifier, so we add ``TAU" as its child  and assign $true$ to the $hasMod$ variable.
    We then reach Line \ref{line:hasMod} where the condition holds and we make a recursive call
    to the children of ``of", {\textit i.e.}, the node $object$ is now ``authors".
    Again we consider all cases until reaching the fourth (Line \ref{line:verbMod}). The condition holds
    since the node ``published" is a verb modifier, thus we replace ``authors" with ``Tova M.", mapped to it. Then, we make a recursive call for all children of ``published"
    since the condition in Line \ref{line:hasMod} holds.
    The nodes ``who" and ``papers" are leaves so they satisfy the condition in Line \ref{line:leaf}.
    Only ``papers" has a value mapped to it, so it is replaced by this value (``OASSIS..."). However,
    the nodes ``after" and ``in" are modifiers so when the algorithm is invoked with $object = \text{``after" (``in")}$,
    the second condition holds (Line \ref{line:mod}) and we replace the subtree of these nodes with
    the node mapped to their child (in the case of ``after" it is ``2014" and in the case of ``in" it is ``SIGMOD") and
    we attach the node ``in" as the parent of the node, in both cases as it is the suitable word for equality for years
    and locations. We obtain a tree representation of the answer (Fig. \ref{fig:organswertree}).

\end{example}

\subsection{From Answer Tree to an Answer Sentence}\label{sec:sentence}

So far we have augmented the NL query dependency tree to obtain the
dependency tree of the answer. The last step is to translate this
tree to a sentence. To this end, we recall that the original query,
in the form of a sentence, was translated by  \nalir\ to the NL
query dependency tree. To translate the dependency tree to a
sentence, we essentially ``revert" this process, further using the
mapping of NL query dependency tree nodes to (sets of) nodes of the
answer. When generating the sentence, we have two different scenarios; when a word or phrase in 
the original dependency tree was replaced by the value to which it was mapped to, 
we replace the word/phrase in the NL query with the value mapped to it. Otherwise, the 
value mapped to it was added as its child, and in this case we add it either before or 
after the mapped word/phrase according to its $POS$ with the appropriate connecting word taken from a stored table.

\begin{example}\label{ex:treeToSentence}
    Converting the answer tree in Figure \ref{fig:organswertree} to a sentence is done
    by replacing the words of the NL query with the values mapped to them,
    {\textit e.g.}, the word ``authors" in the NL query (Figure \ref{fig:nlcq}) is replaced by
    ``Tova M." and the word ``papers" is replaced by ``OASSIS...".
    The word ``organization" is not replaced (as it remains in the answer tree) but rather
    the words ``TAU is the" are added prior to it, since its $POS$ is not a verb and its $REL$
    is a modifier.
    Completing this process, we obtain the answer shown in Figure \ref{fig:single}.

\end{example}

\subsection{Logical Operators}\label{sec:logicalops}

Logical operators (and, or) and the part of the NL query
they relate to will be converted by \nalir\ to a logical predicate
which will be mapped by the assignment to {\em one value} that
satisfies the logical statement (we consider here logical operators that are compiled by \nalir\ into a single CQ, the UCQ case is considered in Section \ref{sec:ucq}). 
To handle these parts of the query,
we augment Algorithm \ref{algo:ansTree} as follows: immediately
following the first case (before the current Line 5), we add a
condition checking whether the node $object$ has a logical operator
(``and" or ``or") as a child. If so, we call Procedure
\ref{algo:logicOps} with the trees $T_Q$ and $T_A$, the logical
operator node as $u$, the \trans\ $\tau$ and the assignment
$\alpha$. The procedure initializes a set $S$ to store the nodes
whose subtree needs to be replaced by the value given to the logical
predicate (Line \ref{line:emp}). Procedure \ref{algo:logicOps} first
locates all nodes in $T_Q$ that were mapped by the \trans\ to the
same query variable as the sibling of the logical operator (denoted
by $u$). Then, it removes the subtrees rooted at each of their
parents  (Line \ref{line:rem}), adds the value (denoted by $val$)
from the database mapped to all of them in the same level as their
parents (Line \ref{line:ins}), and finally, the suitable word for
equality is added as the parent of $val$ in the tree by the
procedure $AddParent$ (Line \ref{line:addparent}).

\IncMargin{1em}
\begin{procedure}[!htb]

    \SetKwInOut{Input}{input}
    \SetKwInOut{Output}{output}
    \LinesNumbered
    \Input{A dependency tree $T_Q$, $T_A$, $u \in V_{T_A}$, \trans\ $\tau$ and an assignment $\alpha$}
    \BlankLine
	$w \gets parent_{T_Q} (u)$\;
	$S \gets \{w\}$\; \label{line:emp}
	 $var \gets \tau(children_{T_A}(w) \setminus u)$\;\label{line:type}
	 $val \gets \alpha(\tau(children_{T_A}(w) \setminus u))$\;\label{line:val}
	 \For {$z \in siblings_{T_A}(w)$}
	 {\label{line:sib}
	     \If{$z$ has child mapped by $\tau$ to $var$}
	     {\label{line:ztype}
	         $S.Insert(z)$\;\label{line:insz}
	     }
	 }
	$parent_{T_A}(w).children_{T_A}().Remove(S)$\;\label{line:rem}
	$parent_{T_A}(w).children_{T_A}().Insert(val)$\;\label{line:ins}
	$AddParent(T_A, val)$ \; \label{line:addparent}
    \caption{HandleLogicalOps()} \label{algo:logicOps}
\end{procedure} \DecMargin{1em}

\section{Factorized Explanations for Multiple Assignments} \label{sec:multiple}

In the previous section we have considered the case where the provenance consists of a single assignment. In general, as illustrated in Section \ref{sec:prelim}, it may include multiple 
assignments. This is the case already for Conjunctive Queries, as illustrated in Section 2.  We next generalize the construction to account
for multiple assignments. Note that a na\"{\i}ve solution in this
respect is to generate a sentence for each individual assignment and
concatenate the resulting sentences. However, already for the
small-scale example presented here, this would result in a long and
unreadable answer (recall Figure \ref{fig:orgprov} consisting of six
assignments). Instead, we propose two solutions: the first based on
the idea of provenance factorization \cite{Olteanu2012,Chapman2008},
and the second (in the following section) leveraging factorization to provide a summarized
form.

\begin{figure}
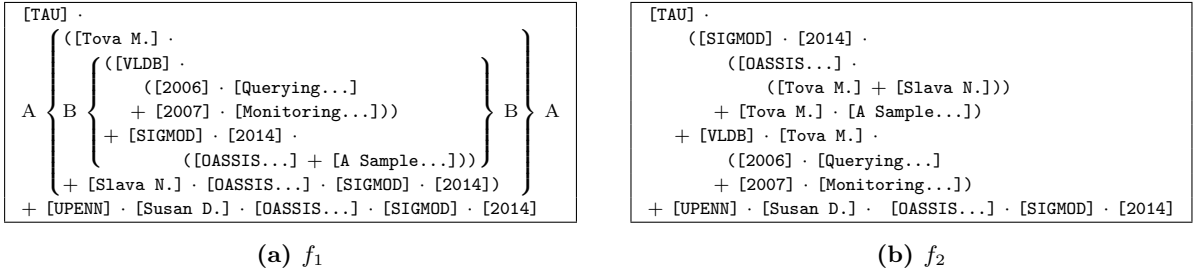

    \centering
    \begin{subfigure}[b]{.5\textwidth}
        \centering
        \scriptsize
        \begin{tabular}{|l|}
            \hline
            \verb"[TAU]" $\cdot$\\
            A $\left\{
            \begin{tabular}{@{}l@{}}
            \verb"([Tova M.]" $\cdot$\\
            B $\left\{
            \begin{tabular}{@{}l@{}}
                \verb"([VLDB]" $\cdot$\\
                \verb"    ([2006]" $\cdot$ \verb"[Querying...]"\\
                \verb"  " $+$ \verb"[2007]" $\cdot$ \verb"[Monitoring...]))"\\
                $+$ \verb"[SIGMOD]" $\cdot$ \verb"[2014]" $\cdot$\\
                \verb"        ([OASSIS...]" $+$ \verb"[A Sample...]))"
            \end{tabular}
            \right \}$ B
            \\
            $+$ \verb"[Slava N.]" $\cdot$ \verb"[OASSIS...]" $\cdot$ \verb"[SIGMOD]" $\cdot$ \verb"[2014])"
            \end{tabular}
            \right \}$ A
            \\
            $+$ \verb"[UPENN]" $\cdot$ \verb"[Susan D.]" $\cdot$ \verb"[OASSIS...]" $\cdot$ \verb"[SIGMOD]" $\cdot$ \verb"[2014]"\\
            \hline
        \end{tabular}
        \caption{$f_1$}
    \end{subfigure}\hfill%
    \begin{subfigure}[b]{.5\textwidth}
        \centering
        \scriptsize
        \begin{tabular}{|l|}
            \hline
            \verb"[TAU]" $\cdot$ \\
            \verb"    ([SIGMOD]" $\cdot$  \verb"[2014]" $\cdot$\\
            \verb"        ([OASSIS...]" $\cdot$ \\
            \verb"            ([Tova M.]" $+$ \verb"[Slava N.]))"\\
            \verb"      "  $+$ \verb"[Tova M.]" $\cdot$ \verb"[A Sample...])"\\
            \verb"  "  $+$ \verb"[VLDB]" $\cdot$ \verb"[Tova M.]" $\cdot$ \\
            \verb"        ([2006]"  $\cdot$ \verb"[Querying...]"\\
            \verb"      " $+$ \verb"[2007]" $\cdot$ \verb"[Monitoring...])"\\
            $+$ \verb"[UPENN]" $\cdot$ \verb"[Susan D.]" $\cdot$ \verb" [OASSIS...]" $\cdot$ \verb"[SIGMOD]" $\cdot$ \verb"[2014]"\\
            \hline
        \end{tabular}
        \caption{$f_2$}
    \end{subfigure}

    \caption{Provenance Factorizations} \label{fig:provfact}
\end{figure}

\subsection{NL-Oriented Factorization}
Provenance size and complexity is a  known and well-studied issue, and various solutions were presented to reduce it \cite{vldb15deutch,factorize}.
Observing that different assignments in the provenance expression typically share significant parts,
one prominent approach \cite{factorize,Olteanu2012,Chapman2008} suggests using algebraic factorization.
The idea is to regard the provenance as a polynomial (see Figure \ref{fig:orgprov})
and use distributivity to represent it in a more succinct way.
For instance, the expression $x \cdot y + x \cdot z$ can be factorized to
the equivalent expression $x \cdot (y + z)$.

The main purpose of classical provenance factorization, as in algebraic factorization, is to reduce the size of the provenance by removing duplicate records and nodes. In our setting, different considerations come into play, as we shall show.

We start by defining the notion of factorization in a standard way
(see {\textit e.g.} \cite{Olteanu2012,elbassioni2011}).

\begin{definition}
Let $P$ be a provenance expression. We say that an expression $f$ is
a factorization of $P$ if $f$ may be obtained from $P$ through
(repeated) use of some of the following axioms: distributivity of
summation over multiplication, associativity and commutativity of
both summation and multiplication.
\end{definition}

\begin{example} \label{example:factexample}

    Re-consider the provenance expression in Figure \ref{fig:orgprov}. Two possible factorizations are shown in Figure
    \ref{fig:provfact}, keeping only the values and omitting the
    variable names for brevity (ignore the A,B brackets for now).
    In both cases, the idea is to avoid repetitions in the provenance expression, by taking out a common factor that appears in multiple summands.
    Different choices of which common factor to take out lead to different factorizations.
\end{example}

How do we measure whether a possible factorization is
suitable/preferable to others? Standard desiderata
\cite{Olteanu2012,elbassioni2011} are that it should be short or
that the maximal number of appearances of an atom is minimal. On the
other hand, we factorize here as a step towards generating an NL
answer; to this end, it will be highly useful if the {\em (partial)
order of nesting of value annotations in the factorization is
consistent the (partial) order of corresponding words in the NL
query}. We will next formalize this intuition as a constraint over
factorizations. We start by defining a partial order on nodes in a
dependency tree:

\begin{definition}\label{def:descendant}
    Given an dependency tree $T$,
    we define $\leq_T$ as the descendant partial order of nodes in $T$:
    for each two nodes, $x, y \in V(T)$, we say that $x \leq_T y$ if $x$ is a descendant of $y$ in $T$.
\end{definition}

\begin{example}
    \label{ex:deporder}
In our running example (Figure \ref{fig:parseMap}) it holds in particular that
$authors \leq organization$, $2005 \leq authors$, $conferences \leq authors$ and $papers \leq authors$, but $papers$, $2005$ and $conferences$ are incomparable. 
\end{example}

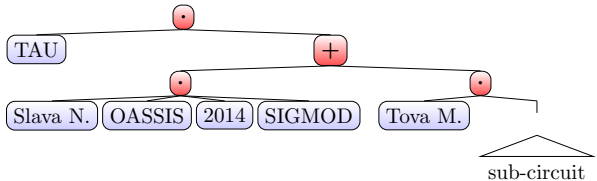
\begin{figure}[htb]
    \centering
    \begin{tikzpicture}[scale=.8,
    level 1/.style={level distance=0.5cm, sibling distance=-10mm},
    level 2/.style={sibling distance=3mm}, level distance=0.6cm, 
    level 3/.style={level distance=0.5cm},
    level 4/.style={level distance=0.2cm, sibling distance=-10mm},
    snode/.style = {shape=rectangle, rounded corners, draw, align=center, top color=white, bottom color=blue!20},
    logic/.style = {shape=rectangle, rounded corners, draw, align=center, top color=white, bottom color=red!70}]
    \Tree
    [.\node[logic] {$\boldsymbol{\cdot}$}; \node[snode] {TAU};
    [.\node[logic] {$\boldsymbol{+}$};
    [.\node[logic] {$\boldsymbol{\cdot}$}; \node[snode] {Slava N.}; \node[snode] {OASSIS}; \node[snode] {2014}; \node[snode] {SIGMOD};]
    [.\node[logic] {$\boldsymbol{\cdot}$}; \node[snode] {Tova M.}; [\qroof{sub-circuit}. ] ]
    ]
    ]
    \end{tikzpicture}
    \caption{Sub-Circuit of $f_1$} \label{fig:circuit}
\end{figure}

Next we define a partial order over elements of a factorization, intuitively based on their nesting depth. To this end, we first consider the {\em circuit form} \cite{Brgisser2010} of a given factorization:

\begin{example}

    Consider the partial circuit of $f_1$ in Figure \ref{fig:circuit}.
    The root, $\cdot$, has two children; the left child is the leaf ``TAU" and
    the right is a $+$ child whose subtree includes the part that is
    ``deeper" than ``TAU".
\end{example}

 Given a factorization $f$ and an element $n$ in it, we denote by $level_f(n)$ 
 the distance of the node $n$ from the root of the circuit induced by $f$ multiplied by $(-1)$. 
 Intuitively, $level_f(n)$ is bigger for a node $n$ closer to the circuit root.

Our goal here is to define the correspondence between the level of
each node in the circuit and the level of its ``source" node in the
NL query's dependency tree (note that each node in the query
corresponds to possibly many nodes in the circuit: all values
assigned to the variable in the different assignments). In the
following definition we will omit the database instance for brevity
and denote the provenance obtained for a query with dependency tree
$T$ by $prov_T$. Recall that \trans\ maps the nodes of the
dependency tree to the query variables and the assignment maps these
variables to values from the database (Definitions
\ref{def:translation}, \ref{def:assignment}, respectively).

\begin{definition} \label{def:nice}
    Let $T$ be a query dependency tree, let $prov_T$ be a provenance expression, let $f$ be a factorization of $prov_T$,
    let $\tau$ be a \trans\
    and let $\{\alpha_1,...\alpha_n\}$ be the set of assignments to the query. For each two nodes $x, y$ in $T$ we say that $x \leq_f y$
    if $\forall i \in [n] : level_f(\alpha_i(\tau(x))) \leq
    level_f(\alpha_i(\tau(y)))$.

    We say that $f$ is $T$-compatible if each pair of nodes $x \neq y \in V(T)$ that satisfy $x \leq_T y$ also satisfy that $ x \leq_f y$.
\end{definition}

Essentially, $T$-compatibility means that the partial order of
nesting between values, for each individual assignment, must be
consistent the partial order defined by the structure of the
question. Note that the compatibility requirement imposes
constraints on the factorization, but it is in general far from
dictating the factorization, since the order $x \leq_T y$ is only
partial -- and there is no constraint on the order of each two
provenance nodes whose ``origins" in the query are unordered. Among
the $T$-compatible factorizations, we will prefer shorter ones.

\begin{definition}
    Let $T$ be an NL query dependency tree and let $prov_T$ be a provenance expression for the answer.
    We say that a factorization $f$ of $prov_T$ is {\em optimal} if $f$ is $T$-compatible and there is no $T$-compatible factorization $f'$ of
    $prov_T$ such that $\mid f' \mid < \mid f \mid$ ($\mid f \mid$ is the length of $f$).
\end{definition}

The following example shows that the $T$-compatibility constraint
still allows much freedom in constructing the factorization. In
particular, different choices can (and sometimes should, to achieve
minimal size) be made for different sub-expressions, 
including ones leading to different answers and ones leading to the same answer through different assignments.

\begin{example}
    \label{ex:factorizeimportant}
    Recall the partial order $\leq_{T}$ imposed by our running example query, shown in part in Example \ref{ex:deporder}. It implies that in every compatible factorization,
    the organization name must reside at the highest level, and indeed $TAU$ was ``pulled out" first in Figure \ref{fig:circuit};
    similarly the author name must be pulled out next. In contrast, since the query nodes corresponding to title, year and conference name are unordered, we may, {\em within a single factorization}, factor out {\textit e.g.} the year in one part of the factorization and the conference name in another one. As an example, Tova M. has two papers published in VLDB (``Querying...'' and ``Monitoring'') so factorizing based on VLDB would be the best choice for that part.
    On the other hand, suppose that Slava N. had two paper published in $2014$; then we could factorize them based on
    $2014$. The factorization could, in that case, look like the
    following (where the parts taken out for Tova and Slava are shown
    in bold):

    \small{
    \begin{tabular}{l}
            \verb"[TAU]" $\cdot$\\
            \begin{tabular}{@{}l@{}}
            \verb"([Tova M.]" $\cdot$\\
            \begin{tabular}{@{}l@{}}
                \verb"("\textbf{[VLDB]} $\cdot$\\
                \verb"    ([2006]" $\cdot$ \verb"[Querying...]"\\
                \verb"  " $+$ \verb"[2007]" $\cdot$ \verb"[Monitoring...]))"\\
                $+$ \textbf{[SIGMOD]} $\cdot$ \verb"[2014]" $\cdot$\\
                \verb"        ([OASSIS...]" $+$ \verb"[A Sample...]))"
            \end{tabular}
            \\
            $+$
            \verb"([Slava N.]" $\cdot$\\
            \begin{tabular}{@{}l@{}}
                \verb"  ("\textbf{[2014]} $\cdot$\\
                \verb"    ([SIGMOD]" $\cdot$ \verb"[OASSIS...]"\\
                \verb"  " $+$ \verb"[VLDB]" $\cdot$ \verb"[Ontology...])))"
            \end{tabular}
            \end{tabular}

        \end{tabular}
        }
    
\end{example}

The following example shows that in some cases, requiring
compatibility can come at the cost of compactness.

As a sanity check, note that the identity factorization that simply
keeps the provenance intact is $T$-compatible. Further, $T$-compatible factorizations are factorizations that keep the answer
(the $object$ node in Figure \ref{fig:abs}) at the start of the sentence and only then refer to the provenance.
When many answers and assignments are involved, it is thus possible to obtain a $T$-compatible factorization by factorizing each answer with its provenance by itself and then combining all of them under a joint root.

\begin{example}\label{ex:tcomp}
    Consider the query tree $T$ depicted in Figure \ref{fig:parseMap} and the factorizations
    $prov_T$ (the identity factorization) depicted in Figure \ref{fig:orgprov}, $f_1$, $f_2$ presented in Figure \ref{fig:provfact}.
    $prov_T$ is of length $30$ and is $5$-readable, {\textit i.e.}, the maximal number of appearances of a single variable is $5$ (see \cite{elbassioni2011}).
    $f_1$ is of length $20$, while the length of $f_2$ is only $19$.
    In addition, both $f_1$ and $f_2$ are $3$-readable.
     Based on those measurements
    $f_2$ seems to be the best factorization, yet
    $f_1$ is $T$-compatible with the question and $f_2$ is not. For example, $conferences \leq_T authors$ but ``SIGMOD" appears higher than ``Tova M." in $f_2$. Choosing a $T$-compatible factorization in  $f_1$ will lead (as shown below) to an answer whose structure resembles that of the question, and thus translates to a more coherent and fitting NL answer.
\end{example}

As mentioned above, the identity factorization is always $T$-compatible, so we
are guaranteed at least one optimal factorization (but it is not
necessarily unique). We next study the problem of computing such a 
factorization.

\subsection{Computing Factorizations}\label{subsec:greedy}

Recall that our notion of compatibility restricts the factorization so that its structure resembles that of the question.
Without this constraint, finding shortest factorizations is coNP-hard in the size of the provenance ({\textit i.e.} a boolean expression) \cite{Hemaspaandra}.
The compatibility constraint does not reduce the complexity since it only restricts choices relevant to part of the expression,
while allowing freedom for arbitrarily many other elements of the provenance. Also recall (Example \ref{ex:factorizeimportant})
that the choice of which element to ``pull-out" needs in general to be done separately for each part of the provenance so as to optimize its size (which is the reason for the hardness in \cite{Hemaspaandra} as well).
In general, obtaining the minimum size $T$-compatible factorization of $prov_T$ is coNP-hard by a reduction from \cite{Hemaspaandra}.

{\bf Greedy Algorithm.} Despite this result, the constraint of
compatibility does help in practice, in that we can avoid examining
choices that violate it. For choices that do maintain compatibility, we devise a simple
algorithm that chooses greedily among them. More concretely, the
input to Algorithm \ref{algo:greedyfactoriation} is the query tree
$T_{Q}$ (with its partial order $\leq_{T_Q}$), and the provenance
$prov_{T_Q}$. The algorithm output is a $T_Q$-compatible
factorization $f$. Starting from $prov$, the progress of the
algorithm is made in steps, where at each step, the algorithm
traverses the circuit induced by $prov$ in a BFS manner from top to
bottom and takes out a variable that would lead to a minimal
expression out of the valid options that keep the current
factorization $T$-compatible. Naturally, the algorithm does not
guarantee an optimal factorization (in terms of length), but
performs well in practice (see Section \ref{sec:experiments}).

In more detail, we start by choosing the largest nodes according to $\leq_{T_Q}$ which have not been processed yet (Line \ref{line:nodes}).
Afterwards, we sort the corresponding variables in a greedy manner based on the number of appearances of
each variable in the expression using the procedure $sortByFrequentVars$ (Line \ref{line:sort}). 
In Lines \ref{line:handleVar}--\ref{line:extract}, we iterate over the sorted variables and extract them from their sub-expressions.
This is done while preserving the $\leq_{T_Q}$ order with the larger nodes, thus ensuring that the factorization will remain $T_Q$-compatible.
We then add all the newly processed nodes to the set\\$Processed$ which contains all nodes that have already been processed (Line \ref{line:add}).
Lastly, we check whether there are no more nodes to be processed, {\textit i.e.}, if the set $Processed$ includes all
the nodes of $T_Q$ (denoted $V(T_Q)$, see the condition in Line \ref{line:doneFact}).
If the answer is ``yes", we return the factorization.
Otherwise, we make a recursive call. In each such call, the set $Processed$ becomes larger until the condition in Line \ref{line:doneFact} holds.

\IncMargin{1em}
\begin{algorithm}[!htb]
    \SetKwFunction{GreedyFactorization}{GreedyFactorization}
    \SetKwInOut{Input}{input}\SetKwInOut{Output}{output}
    \LinesNumbered

    \Input{$T_Q$ - the query tree, $\leq_{T_Q}$ - the query partial order, $prov$ - the provenance, $\tau, \alpha$ - \trans\ and assignment from nodes in $T_Q$ to provenance variables,  $Processed$ -
    subset of nodes from $V(T_Q)$ which were already processed (initially, $\emptyset$)}
    \Output{$f$ - $T_Q$-compatible factorization of $prov_{T_Q}$} \BlankLine

    $f \gets prov$\; \label{line:copy}
    $Frontier \gets \{x \in V(T_Q) | \forall (y \in V(T_Q) \setminus Processed) \;\; s.t.~ x  \not \leq_{T_Q} y\}$\;\label{line:nodes}
    $vars \gets sortByFrequentVars(\{\alpha(\tau(x)) | x \in Frontier \}, f)$\;  \label{line:sort} 
    \ForEach {$var \in vars$}
    {\label{line:handleVar}
        Take out $var$ from sub-expressions in $f$ not including variables from $\{x | \exists y \in Processed : x = \alpha(\tau(y))\}$\; \label{line:extract}
    }
    $Processed \gets Processed \cup Frontier$\; \label{line:add}
    \If {$|Processed| = |V(T_Q)|$}
    {\label{line:doneFact}
        \Return $f$\;
    }
    \Else
    {\label{line:recFact}
        \Return $GreedyFactorization(T_Q, f, \tau, \alpha, Processed)$\;
    }

    \caption{GreedyFactorization}
    \label{algo:greedyfactoriation}
\end{algorithm} 
\DecMargin{1em}

\begin{example}\label{ex:factalgo}

    Consider the query tree $T_Q$ depicted in Figure \ref{fig:parseMap}, and provenance $prov$ in Figure \ref{fig:orgprov}.
    As explained above, the largest node according to $\leq_{T_Q}$ is $organization$, hence ``TAU" will be taken out from the brackets multiplying all summands that  contain it. Afterwards, the next node
    according to the order relation will be $author$, therefore we group by author, taking out ``Tova M.", ``Slava N." etc. The following choice (between conference, year and paper name) is then done greedily {\em for each author}, based on its number of occurrences. For instance, $VLDB$ appears twice for $Tova. M.$ whereas each paper title and year appears only once; so it will be pulled out.
    The polynomial $[Slava N.] \cdot [OASSIS...] \cdot [SIGMOD] \cdot [2014]$ will remain unfactorized as all values appear once.  Eventually, the algorithm will return the factorization $f_1$ depicted in Figure \ref{fig:provfact}, which is $T_Q$-compatible and much shorter than the initial provenance expression.
\end{example}

\paragraph{Complexity} 
Denote the provenance size by $n$. The algorithm complexity is
$O(n^{2} \cdot \log n)$: at each recursive call, we sort all nodes
in $O(n \cdot \log n)$ (Line \ref{line:sort}) and the we handle (in
$Frontier$) at least one node (in the case of a chain graph) or
more. Hence, in the worst case we would have $n$ recursive calls,
each one costing $O(n \cdot \log n)$.

\paragraph{Optimization}
Since $T$-compatible factorizations keep the answer
(the $object$ node in Figure \ref{fig:abs}) at the start of the sentence,
we can utilize the {\em abstract factorization structure} for one answer
in the factorization of all other answers. For this, we need to augment
Algorithm \ref{algo:greedyfactoriation} in the following manner.
First, only the monomials that contain the first answer will be factorized using Algorithm \ref{algo:greedyfactoriation}.
Then, an abstract factorization structure $f_a$ can be inferred from this factorization by
replacing some of the values with the variables mapped to them. The values
that are replaced with variables are those that have a clear hierarchy between them in $T_Q$
while values that were mapped to words in the same level of $T_Q$ are not part of the
abstract factorization structure as the hierarchy between them may vary based on the nature of
the assignments each results has. Namely, if $x \neq y \in V(T_Q)$ satisfy $x \leq_{T_Q} y$,
the variables that $x$ and $y$ are mapped to will be part of $f_a$
and will hold $var(x) \leq_{f_a} var(y)$ where $var(x)$ is the variable $x$ is mapped to.
Intuitively, the circuit induced by $f_a$ maintains the partial order of nodes in $T_Q$.
Finally, given the provenance polynomial of another answer,
we replace the variables in $f_a$ with the corresponding constants and
greedily factorize only the parts of the polynomial that are not included in $f_a$.

\subsection{Factorization to Answer Tree}
The final step is to turn the obtained factorization into an NL
answer. Similarly to the case of a single assignment (Section
\ref{sec:single}), we devise a recursive algorithm that leverages
the mappings and assignments to convert the query dependency tree
into an answer tree. Intuitively, we follow the structure of a
single answer, replacing each node there by either a single node,
standing for a single word of the factorized expression, or by a
recursively generated tree, standing for some brackets (sub-circuit)
in the factorized expression.

In more detail, the algorithm operates as follows. We iterate over
the children of $root$ (the root of the current sub-circuit),
distinguishing between two cases. First, for each leaf child, $p$,
we first (Line \ref{line:assign}) assign to $val$ the database value
corresponding to the first element of $p$ under the assignment
$\alpha$ (recall that $p$ is a pair (variable,value)). We then
lookup the node containing the value mapped to $p$'s variable in the
answer tree $T_A$ and change its value to $val$ in Lines
\ref{line:lookup}, \ref{line:repNode} (the value of $p$). Finally,
in Line \ref{line:rearrange} we reorder nodes in the same level
according to their order in the factorization (so that we output a
semantically correct NL answer).
Second, for each non-leaf child, the algorithm performs a
recursive call in which the factorized answer subtree is computed (Line \ref{line:rec}).
Afterwards, the set containing the nodes of the resulting subtree aside from
the nodes of $T_A$ are attached to $T_{F}$ under the node corresponding to their LCA in $T_{F}$ (Lines \ref{line:recNodes} -- \ref{line:attach}).
In this process, we attach the new nodes that were placed lower in the circuit in
the most suitable place for them semantically (based on $T_A$), while also maintaining the structure of the factorization.

\IncMargin{1em}
\begin{algorithm}[!htb]
    \SetKwFunction{ComputeFactAnswerTree}{ComputeFactAnswerTree}
    \SetKwInOut{Input}{input}\SetKwInOut{Output}{output}
    \LinesNumbered
    \Input{$\alpha$ - an assignment to the NL query,
        $T_A$ - answer dependency tree based on $\alpha$,
        $root$ - the root of the circuit induced by the factorized provenance
        }
    \Output{$T_{F}$ - tree of the factorized answer} \BlankLine

    $T_{F} \gets copy(T_A)$\;

    \ForEach {$p \in children_f(root)$}
    {\label{line:iter}
    \If {$p$ is a leaf}
    {\label{line:externalVars}
        $val \gets \alpha(var(p))$\;\label{line:assign}
        $node \gets Lookup(var(p), \alpha, T_A)$\; \label{line:lookup}
        $ReplaceVal(val, node, T_F)$\;\label{line:repNode}
        $Rearrange(node, T_A, T_F)$\; \label{line:rearrange}
    }
    \Else
    {
        {\label{line:subExp}
            $T^{rec}_{F} = ComputeFactAnswerTree(\alpha, T_A, p)$\;\label{line:rec}
            $RecNodes = V(T^{rec}_{F}) \setminus V(T_A)$\;\label{line:recNodes}
            $parent^{rec}_{F} \gets LCA(recNodes)$\;
            $parent_{F} \gets$ Corresponding node to $parent^{rec}_{F}$ in $T_{F}$\;
            Attach $recNodes$ to $T_{F}$ under the $parent_{F}$\;\label{line:attach}
        }
    }
    }

    \Return $T_{F}$\; \label{line:return}

    \caption{ComputeFactAnswerTree} \label{algo:factorizedTree}
\end{algorithm} \DecMargin{1em}

\begin{example}
\label{ex:factorizations}
    Consider the factorization $f_1$ depicted in Figure \ref{fig:provfact}, and the structure of single assignment
    answer depicted in Figure \ref{fig:organswertree} which was built based on Algorithm \ref{algo:ansTree}.
    Given this input, Algorithm \ref{algo:factorizedTree} will generate an answer tree corresponding to the following sentence:

    \small{
        \begin{tabular} {l}
            \verb"TAU is the organization of"\\
            \verb"    Tova M. who published"\\
            \verb"        in VLDB"\\
            \verb"           'Querying...' in 2006 and"\\
            \verb"           'Monitoring...' in 2007"\\
            \verb"        and in SIGMOD in 2014"\\
            \verb"            'OASSIS...' and  'A sample...'"\\
            \verb"    and Slava N. who published"\\
            \verb"        'OASSIS...' in SIGMOD in 2014."\\
            \verb"UPENN is the organization of Susan D. who published"\\
            \verb"'OASSIS...' in SIGMOD in 2014."\\
        \end{tabular}
    }

        \normalsize
        Note that the query has two results: ``TAU" and ``UPENN". ``UPENN" was produced with a single assignment,
        but there are 5 different assignments producing ``TAU". We now focus on this sub-circuit depicted in Figure \ref{fig:circuit}.
        After initializing $T_{F}$, in Lines \ref{line:externalVars} -- \ref{line:rearrange} the algorithm finds the value $TAU$ and the node corresponding to it in $T_A$ (which originally contained the variable $organization$).
        It then copies this node to $T_F$ and assigns it the value ``TAU".
        Next the algorithm handles the $+$ node with a recursive call in Line \ref{line:rec}.
        This node has the two sub-circuits rooted at the two $\cdot$ nodes
        (Line \ref{line:subExp}); one containing $[authors, Tova M.]$ and the other $[authors, Slava N.]$.
        When traversing the sub-circuit containing ``Slava N.", the algorithm simply copies the subtree rooted at the $authors$ node with the values from the circuit
        and arranges the nodes in the same order as the corresponding variable nodes were in $T_A$ (Line \ref{line:rearrange}) as they are all leaves on the same level.
        Those values will be attached under the LCA ``of" (Lines \ref{line:rec} -- \ref{line:attach}).
        The sub-circuit of ``Tova M." also has nested sub-circuits. Although the node $paper$ appears before the nodes $year$ and $conference$ in the answer tree structure,
        the algorithm identifies that $f_1$ extracted the variables ``VLDB", ``SIGMOD" and ``2014", so it changes their location so that they appear earlier in the final answer tree.
        Finally, recursive calls are made with the sub-circuit containing $[authors, Tova M.]$.

        Intuitively,  ``of" is indeed the root of a sub-tree specifying the authors in an institution in the structure of our answers.

\end{example}

\section{From Factorized to Summarized Answers}\label{subsec:summ}

So far we have proposed a solution that factorizes multiple assignments,
leading to a more concise answer. When there are many assignments and/or the assignments involve multiple distinct values, even an optimal factorized representation may be too long and convoluted for users to
follow.

\begin{example}
Reconsider Example \ref{ex:factorizations}; if there are\\many authors from TAU then even the compact representation of the result could be very long.
In such cases we need to summarize the provenance in some way that will preserve the ``essence'' of all assignments without
actually specifying them, for instance by providing only the number of authors/papers for each institution.
\end{example}

To this end, we employ {\em summarization}, as follows. First, we note that a key to summarization is understanding which parts of the provenance may be grouped together. For that, we use again the mapping from nodes to query variables: we say that two nodes are of the same {\em type} if both were mapped to the same query variable. Now, let $n$
be a node in the circuit form of a given factorization $f$. A summarization of the sub-circuit of $n$ is obtained in two steps. First, we group the descendants of $n$ according to their type. Then, we summarize each group separately. The latter is done in our implementation simply by either counting the number of distinct values in the group or by computing their range if the values are numeric. In general, one can easily adapt the solution to apply additional user-defined ``summarization functions" such as ``greater / smaller than X" (for numerical values) or ``in continent Y" for countries.

\begin{figure}
    \centering
    \scriptsize{
        \begin{tabular}{|l|}
            \hline
            (A) \verb"[TAU]" $\cdot$ {\em Size}\verb"([Tova M.],[Slava N.])" $\cdot$ {\em Size}\verb"([VLDB],[SIGMOD])" $\cdot$ \\
            \verb"     " {\em Size}\verb"([Querying...],[Monitoring...],"\\
            \verb"       [OASSIS...],[A Sample...])" $\cdot$ {\em Range}\verb"([2006],[2007],[2014])"\\
            (B) \verb"[TAU]"$\cdot$\verb"("\\
            \verb"     [Tova M.]" $\cdot$ \\
            \verb"      " {\em Size}\verb"([VLDB],[SIGMOD])" $\cdot$ \\
            \verb"      " {\em Size}\verb"([Querying...],[Monitoring...]," \\
            \verb"       [OASSIS...],[A Sample...])" $\cdot$ {\em Range}\verb"([2006],[2007],[2014])"\\
            \verb"     [Slava N.]" $\cdot$ \verb"[OASSIS...]" $\cdot$ \verb"[SIGMOD]" $\cdot$ \verb"[2014])" \\
            \hline
        \end{tabular}
    }
    \caption{Summarized Factorizations} \label{fig:factsummary}
\end{figure}

\begin{figure}
    \centering
    \small{
        \begin{tabular}{|l|}
            \hline
            (A) \verb"TAU is the organization of 2 authors who published"\\
            \verb"4 papers in 2 conferences in 2006 - 2014."\\
            (B) \verb"TAU is the organization of Tova M. who published"\\
            \verb"4 papers in 2 conferences in 2006 - 2014 and Slava N."\\
            \verb"who published 'OASSIS...' in SIGMOD in 2014."\\
            \hline
        \end{tabular}
    }
    \caption{Summarized Sentences} \label{fig:sentencesummary}
    \end{figure}

\begin{example}\label{ex:summarize}

    Re-consider the factorization $f_1$ from Figure
    \ref{fig:provfact}. We can summarize it in multiple levels: the highest level of authors (summarization ``A"), or the level of papers for each particular author (summarization ``B"), or the level of conferences, etc. Note that if we choose to summarize at some level, we must summarize its entire sub-circuit ({\textit e.g.} if we summarize for Tova. M. at the level of conferences, we cannot specify the papers titles and publication years).

    Figure \ref{fig:factsummary} presents the summarizations of  sub-trees for the ``TAU" answer, where ``size" is a summarization operator that counts the number of distinct values and ``range" is an operator over numeric values, summarizing them as their range. The summarized factorizations are further converted to NL sentences which are shown in Figure \ref{fig:sentencesummary}.
    Summarizing at a higher level results in a shorter but less detailed summarization.
\end{example}

\section{Unions of Conjunctive Queries}\label{sec:ucq}
So far our solution has been limited to Conjunctive Queries, and we next extend it to account for Unions thereof (UCQs). We next describe the necessary augmentations of the algorithms, illustrating them through examples. 
Recall that in the first step, the system takes a natural language query and translates it to a dependency tree, while maintaining the \trans\ mapping. The difference here is that a tree node can now be mapped to several variables. This implies a generalization of Definition \ref{def:translation}:

\begin{definition}\label{def:translationUCQ}
    Given a dependency tree $T=(V,E,L)$ and a UCQ $Q_1, \ldots, Q_m$, a \transu\ is a set of \trans\ $\{\tau_1, \ldots, \tau_m\}$,  where $\tau_i: T \to Q_i$.
\end{definition}

\begin{example}
Consider the NL query ``return the organization of authors who published papers in database conferences before 2005 or after 2015", whose dependency tree is depicted in Figure \ref{fig:logOps}. The ``or" here defines two different CQs (depicted in Figure \ref{fig:union}). Since the two numerical values cannot form a conjunctive condition and thus cannot be compiled into a single boolean condition, \nalir\ translates this NL query into two CQs. 
The two CQs define two different \trans\ that map nodes from the single dependency tree to two different sets of variables. Consider an organization (e.g., TAU) which appears as an answer. It is mapped both to $oname1$ and to $oname2$. Thus, we would like to present the assignments to both queries as explanations. Here, the \transu\ is $\{\tau_1, \tau_2\}$ where $\tau_1$ maps the nodes from the dependency tree to the variables of the first query in Figure \ref{fig:union} and $\tau_2$ maps the nodes from the dependency tree to the variables of the second query. Thus, the \transu\ captures the assignments from both queries. Note that $\tau_1$ differs from $\tau_2$ for some words. Specifically, $\tau_1$ maps the nodes ``before" and ``2005" to $pyear1 < 2005$ and $\tau_2$ maps the nodes ``after" and ``2015" to $pyear2 > 2005$.
\end{example}

We further give an unique integer identifier to each mapped word in the dependency tree as exemplified by the superscript in Figure \ref{fig:logOps} for reasons we explain in the sequel.

\begin{figure}[htb!]
    \centering
    \begin{tikzpicture}[thick, scale=.82,
    ssnode/.style={fill=mygreen,node distance=.1cm,draw,circle, minimum size=.1mm},
    -,shorten >= 2pt,shorten <= 2pt,
    level 1/.style={level distance=0.2cm},
    every node/.style = {shape=rectangle, rounded corners, draw, align=center, top color=white, bottom color=blue!20},
    map/.style={shape=rectangle, rounded corners, draw, align=center, top color=white, bottom color=red!20},
    log/.style={shape=rectangle, rounded corners, draw, align=center, top color=white, bottom color=orange!60},
    level 2/.style={sibling distance=-17mm},
    level 5/.style={sibling distance=1mm},
    level distance=0.8cm
    ]

    \begin{scope}[level 1/.style={level distance=1.0cm}, yshift= 7cm, xshift= -1.7cm]

    \Tree
    [.return
    [.\node[] (a5) {organization$^1$}; {the}
    [.\node[] (z) {of};
    [.\node[] (a1) {authors$^2$};
    [.{published} {who} 
    [.\node[] (a2) {papers$^3$}; ]
    [.{before} [.\node[] (a4) {2005$^5$}; ] ]
    [.{after} [.\node[] (a4) {2015$^6$}; ] [.\node[log] (a4) {or}; ] ]
    [.{in}
    [.\node[] (a3) {conferences$^4$};
    [.\node[] (a6) {database}; ]
    ]
    ]
    ]
    ]
    ]
    ]
    ]
    \end{scope}
    \end{tikzpicture}
    \caption{Dependency Tree With ``Or" Condition}\label{fig:logOps}
\end{figure}
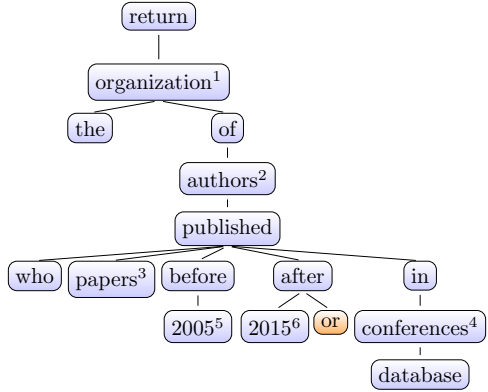

\begin{figure}
  \centering
  \small{
      \begin{tabular}{|l|}
          \hline
          \verb"query(oname1) :- org(oid1, oname1), author(aid1, aname1,"\\
          \verb" oid1), pub(wid1, cid1, ptitle1, pyear1), conf(cid1,"\\
          \verb"cname1), domainConf(cid1, did1), domain(did1, dname1),"\\
          \verb"writes(aid1, wid1), dname1 = 'Databases', pyear1 < 2005"\\
          \hline
          \hline
          \verb"query(oname2) :- org(oid2, oname2), author(aid2, aname2,"\\
          \verb" oid2), pub(wid2, cid2, ptitle2, pyear2), conf(cid2,"\\
          \verb"cname2), domainConf(cid2, did2), domain(did2, dname2),"\\
          \verb"writes(aid2, wid2), dname2 = 'Databases', pyear2 > 2015"\\
          \hline
      \end{tabular}
  }
  \caption{Two CQs from the Same NL Query} \label{fig:union}
\end{figure}

After determining the \transu, we rely on an augmentation of Definition \ref{def:valuelevel}. The following definition essentially generalizes the definition for CQs by also summing the pairs of (word identifier, value) from all the CQs participating in the union.

\begin{definition}
    \label{def:prov_ucq}
    Let $A(Q,D)$ be the set of assignments for a UCQ $\mathcal{Q} = \{Q_1, \ldots,  Q_m\}$ and a database instance $D$, and let $\{\tau_1, \ldots, \tau_m\}$ be the \transu\ of $\mathcal{Q}$. We define the {\em NL value-level provenance} of $\mathcal{Q}$ w.r.t. $D$ as
    $$\sum_{Q_i \in \mathcal{Q}}\sum_{\alpha \in A(Q_i,D)}\Pi_{\{x_i,a_i \mid \alpha(x_i) = a_i\}}(\tau_i^{-1}(x_i),a_i).$$
\end{definition}

\begin{figure}[!htb]
    \centering \scriptsize
    \begin{minipage}{.5\linewidth}
        \centering
        \caption*{Rel. $\text{\textit{org}}$}\label{tbl:organization}
        \begin{tabular}{| c | c | c | c | c | c |}
            \hline oid & oname \\
            \hline 2 & TAU \\
            \hline
        \end{tabular}
        
    \end{minipage}%
    \begin{minipage}{.5\linewidth}
        \centering
        \caption*{Rel. $\text{\textit{author}}$}\label{tbl:author}
        \begin{tabular}{| c | c | c | c | c | c |}
            \hline aid & aname & oid \\
            \hline 4 & Tova M. & 2 \\
            \hline
        \end{tabular}
        
    \end{minipage}

    \begin{minipage}{0.7\linewidth}
        \centering
        \caption*{Rel. $\text{\textit{pub}}$}\label{tbl:publication}
        \begin{tabular}{| c | c | c | c | c | c |}
            \hline wid & cid & ptitle & pyear\\
            \hline 6 & 10 & ``Positive Active XML" & 2004 \\
            \hline 7 & 11 & ``Rudolf..." & 2016 \\
            \hline
        \end{tabular}
        
    \end{minipage}%
    \begin{minipage}{.3\linewidth}
        \centering
        \caption*{Rel. $\text{\textit{writes}}$}\label{tbl:writes}
        \begin{tabular}{| c | c | c | c | c | c |}
            \hline aid & wid\\
            \hline 4 & 6 \\
            \hline 4 & 7 \\
            \hline
        \end{tabular}
        
    \end{minipage}

    \begin{minipage}{.33\linewidth}
        \centering
        \caption*{Rel. $\text{\textit{conf}}$}\label{tbl:conference}
        \begin{tabular}{| c | c | c | c | c | c |}
            \hline cid & cname\\
            \hline 10 & PODS \\
            \hline 11 & VLDB \\
            \hline
        \end{tabular}
        
    \end{minipage}
    \begin{minipage}{.33\linewidth}
        \centering
        \caption*{Rel. $\text{\textit{domainConf}}$}\label{tbl:domainconference}
        \begin{tabular}{| c | c | c | c | c | c |}
            \hline cid & did\\
            \hline 10 & 18 \\
            \hline 11 & 18 \\
            \hline
        \end{tabular}
        
    \end{minipage}%
    \begin{minipage}{.33\linewidth}
        \centering
        \caption*{Rel. $\text{\textit{domain}}$}\label{tbl:domain}
        \begin{tabular}{| c | c | c | c | c | c |}
            \hline did & dname\\
            \hline 18 & Databases \\
            \hline
        \end{tabular}
        
    \end{minipage}

    \caption{DB Instance for Example \ref{ex:provucq}}\label{tbl:dbucq}
\end{figure}

\begin{figure}[]
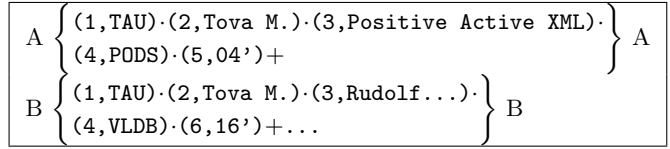

    \begin{center}
    \small{
        \begin{tabular}{|l|}
            \hline
            A $\left\{
            \begin{tabular}{@{}l@{}}
           \verb"(1,TAU)"$\cdot$\verb"(2,Tova M.)"$\cdot$\verb"(3,Positive Active XML)"$\cdot$\\
                                     \verb"(4,PODS)"$\cdot$\verb"(5,04')"$+$ \\
                            \end{tabular}
           \right\}$ A\\
           B $\left\{
           \begin{tabular}{@{}l@{}}
           \verb"(1,TAU)"$\cdot$\verb"(2,Tova M.)"$\cdot$\verb"(3,Rudolf...)"$\cdot$\\
                                     \verb"(4,VLDB)"$\cdot$\verb"(6,16')"$+$\verb"..." \\
                                     \end{tabular}
                                     \right\}$ B\\
            \hline
        \end{tabular}
    }
    \caption{Value-level Provenance for Example \ref{ex:provucq}} \label{fig:orgprovucq}
    \end{center}
\end{figure}

\begin{example}\label{ex:provucq}
Reconsider the UCQ defined by the union of the two CQs depicted in Figure \ref{fig:union} and the database in Figure \ref{tbl:dbucq} with tuples standing for two more publications by the author Tova Milo: ``Positive Active XML" published in PODS in 2004 and ``Rudolf: Interactive Rule Refinement System for Fraud Detection" published in VLDB in 2016. The first of the two summands in Figure \ref{fig:orgprovucq} (in the ``A'' brackets) stands for an assignment to the top query in Figure \ref{fig:union}, while the second summand (in the ``B'' brackets) stands for an assignment for the bottom query. Assignments are represented as multiplication of pairs of the form $(id, val)$ so that $id$ is the unique identifier of a word in the NL query mapped to the variable $var$ in a specific query $Q_i$ that is assigned $val$ in the particular assignment.
\end{example}

We now have a polynomial containing sets of pairs where the first element is the unique word in the NL query and the second is the value from the database mapped to it. This allows us to consider explanations for the same answer {\em regardless of the query from which they originated}.

By replacing the variable name in each pair with the unique word identifier from the NL query, we are able to treat the assignment  of different variable names as relating to the same word or phrase in the NL query. This allows us to factorize the provenance of the different queries in the union in the context of a single NL query to which we will build a single NL answer. Now, we can use the procedure described in Section \ref{sec:multiple} to produce a $T$-compatible factorization and summarization of the provenance. 
The only change needed in Algorithm \ref{algo:factorizedTree} is to replace all nodes that form the logical ``or" condition with the words mapped to them. In our example, replacing the subtrees rooted at ``before" and ``after" with the year from the provenance assignments.

\section{Implementation and Generalizations}\label{sec:implementation}

\subsection{Implementation}

\systemName\ is implemented in JAVA 8, extending \nalir. Its web UI
is built using HTML, CSS and JavaScript. It runs on Windows 8 and
uses MySQL server as its underlying database management system (the
source code is available in \cite{git}). Figure \ref{fig:sysArch}
depicts the system architecture. First, the user enters a query in
Natural Language. This NL sentence is fed to the augmented \nalir\
system which interprets it and generates a formal query. This
includes the following steps: a parser \cite{lrec2006} generates the
dependency tree for the NL query. Then, the nodes of the tree are
mapped to attributes in the tables of the database and to functions,
to form a formal query. In fact, \nalir\ may generate several candidate queries, from which it will choose the one that is ranked highest according to an internal ranking function. We use the highest ranked as the chosen query. As explained, to be able to translate the
results and provenance to NL, \systemName\ stores the mapping from
the nodes of the dependency tree to the query variables. Once a
query has been produced, \systemName\ uses the \selp\ system
\cite{vldb15deutch} to evaluate it while storing the provenance,
keeping track of the mapping of dependency tree nodes to parts of
the provenance. The provenance information is then factorized (see
Algorithm \ref{algo:greedyfactoriation})  and the factorization is
compiled to an NL answer (Algorithm \ref{algo:factorizedTree})
containing
explanations.
Finally, the factorized answer is shown to the user. If the answer
contains excessive details and is too difficult to understand, the
user may choose to view summarizations.

\paragraph{User Interface}
We now discuss the user interface \systemName.
First the user writes a natural language question in the web interface. The question is inputted to the augmented \nalir\ box, converted to an SQL query while storing the mapping from words to variables and evaluated over the database, where the query results . All results are then shown to the user, where each result can be further explored by viewing its natural language provenance, in each of the three forms described earlier: an explanation formed by a single assignment, an explanation which encapsulates all assignments as a factorized representation of the provenance, and a summarized explanation based on the factorization.

\subsection{Replacing the Black Boxes}\label{subsec:arch_extended}

Our solution ``marries", for the first time to our knowledge, two fields: (1) Natural Language Interfaces to Databases, and (2) Data Provenance. For each of these two, we have made choices in our implementation: \nalir{} for the NLIDB, as well as a semiring-like value-level provenance model. We next revisit these choices and discuss alternatives in detail. 

\subsubsection{Alternatives NLIDBs}

As mentioned above, \nalir{} is a prominent interface for querying relational databases in Natural Language. Yet the problem of transforming an NL query into a formal query has been researched extensively, by both the database and NLP communities, and it includes a variety of different approaches for the solution. As the field keeps progressing, the question rises: how flexible is our approach of Natural Language Provenance, with respect to NLIDB development? Namely, if an improved NLIDB is developed, can it be incorporated in our framework?

To address this question, we will analyze our requirements and briefly discuss state-of-the-art algorithms for the problem of translating text to SQL or similar formal languages, reviewing their compatibility to these requirements and consequently the possibility of their binding to \systemName{}. Further in-depth discussion of the works themselves appears in our review of related work in Section 10.

The DB community has been studying \emph{Natural Language Interfaces to Databases} for several decades. Many solutions focus on matching the query parts to the DB schema, and infer the SQL based on this mappings, Obtaining a matching from natural language query to the DB schema in various ways, such as pattern matching, grammar matching, or intermediate representations language (see Section \ref{sec:relatedWork} for more details).

The NLP community has also extensively studied the translation of natural language question to logical representations that query a knowledge base \cite{zettlemoyer2012learning,liang2013learning,berant2014semantic,beltagy2014semantic}.
One of the earliest statistical models for mapping text to SQL was the \precise{} system \cite{popescu2003,popescu2004}; it was able to achieve high precision on specific class of queries that were able to be linked tokens and database values, attributes, and relations. However, \precise{} did not attempt to generate SQL for questions outside this class. 
Later work considered generating queries based on relations extracted by a syntactic parser \cite{giordani2012translating} and applying techniques from logical parsing research \cite{poon2013grounded}.
 Recently there is a flourish of work on generating SQL \cite{yih2015semantic,iyer2017learning,zhong2017seq2sql}, typically  applying Machine Learning methods such as seq2seq networks and reinforcement learning.

\begin{figure}[]
	\begin{subfigure}{.47\textwidth}
		\centering
		\includegraphics[scale = 0.3, trim = 0 0 4mm 4mm]{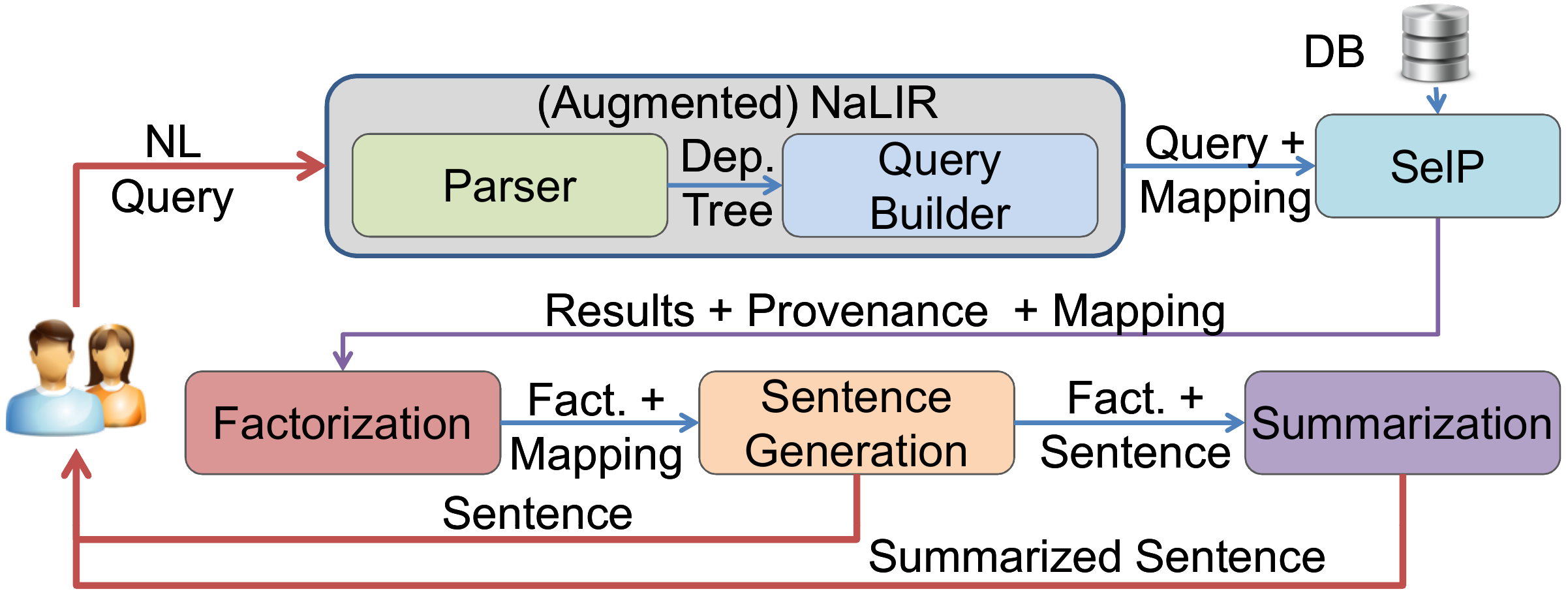}
    \caption{System Architecture}
    \label{fig:sysArch}
	\end{subfigure}%
	\begin{subfigure}{.47\textwidth}
		\centering
		\includegraphics[scale = 0.3, trim = 0 15mm 2mm 0mm]{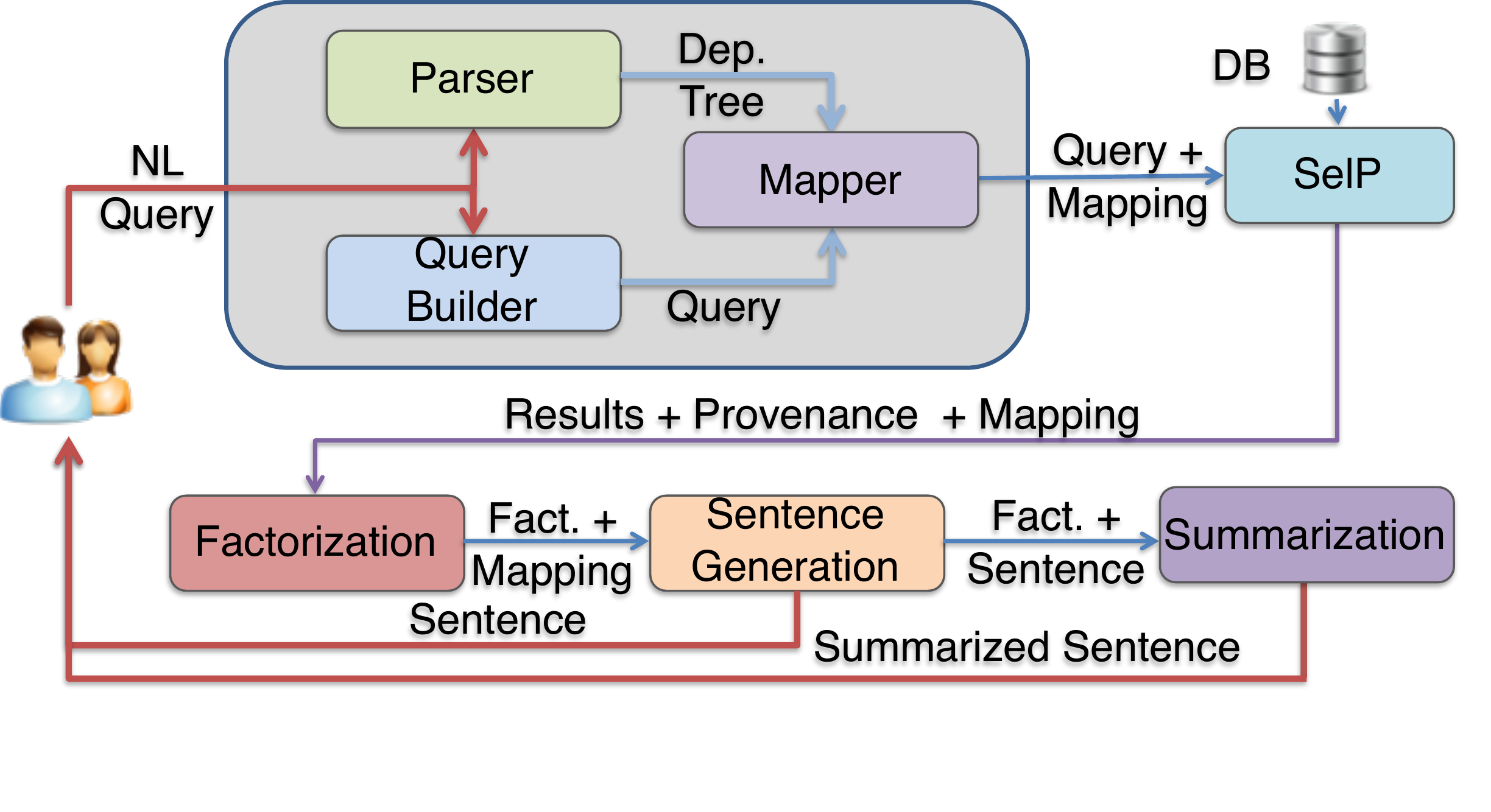}
    \caption{Extended System Architecture}
    \label{fig:sysArchExtendn}
	\end{subfigure}
	\caption{\systemName{} and General Architectures}
	\label{fig:arch}
\end{figure}

\systemName{} architecture, as depicted in Figure \ref{fig:sysArch} and explained previously, is coupled with an augmented version of \nalir{} in the following sense:  we get from  \nalir{} both its translation to a formal query, along with the \emph{dependency-to-query-mapping} $\tau$. These are the two essential factors for the operation of \systemName{}. That is, \systemName{} can use any existing system that transform natural language question to formal query and also return partial mapping from the dependency tree nodes to the query parts. Indeed, many of the other techniques for Natural Language Interface design mentioned above, could be adapted to return $\tau$ and support our requirements. For example \precise{} has a component called ``matcher", that generates mapping from tokens and database values, attributes, and relations. 
However, not all of the methods described above are designed in a way that allows generation of this mapping. E.g., a semantic parser that relies on seq2seq Deep Neural Network may be unable to return this mapping. The DNN would be trained on a large corpus of natural language questions along with their relevant SQL queries, and its objective would be to generalize to new questions. Due to the network complex representation it may be hard to extract the desired mapping.

To this end, we propose an alternative architecture, depicted in Figure \ref{fig:sysArchExtendn}. This architecture does not rely on the query builder to also generate the partial mapping $\tau$ from the dependency tree nodes to the query parts. Instead, we have added an additional block, \emph{Mapper}, that receives as input the dependency tree along with the generated query and outputs the mapping $\tau$. Note that generating the dependency tree may be done using existing tools such as the Stanford Parser, independently of whether the NLIDB generates it (as \nalir{} does) or not (as is the case with semantic parsers). 
\IncMargin{1em}
\begin{algorithm}[!htb]
    \SetKwFunction{Mapper}{Mapper}
    \SetKwInOut{Input}{input}\SetKwInOut{Output}{output}
    \LinesNumbered
    \Input{Dependency tree nodes $V$, \\ Conjunctive Query $Q$, \\ Similarity Threshold $\beta$}
    \Output{Partial Mapping $\tau$} \BlankLine
    
    $G_{vertices} \defeq V \cup VAR(Q)$\;
    $G_{edges} \defeq \emptyset$\;
    \ForEach{$v\in V $}
    {
        \ForEach{$q\in VAR(Q)$} {
            \If{$Sim(v,q) \geq \beta$} {
                $e \defeq \left(v,q\right)$\;
                $e_{weight} \defeq Sim(v,q)$\;
                $G_{edges} \defeq G_{edges} \cup \{e\}$\;
            }
        }
    }
    \Return $MaximalMatching(G)$\;

    \caption{Mapper} \label{algo:mapper}
\end{algorithm} \DecMargin{1em}

We then present Algorithm \ref{algo:mapper} responsible for the mapping generation. The algorithm is similar in spirit to the default mapping algorithms of \nalir{} and \precise{}, but could be used as a stand-alone component without these systems. It generates a bipartite graph, with the dependency tree nodes at one side, and the query parts in the second side. For each pair the algorithm calculates a similarity between the two, and in case they are similar enough (similarity is higher than the input constant $\beta$) an edge will be generated with the corresponding weight. Eventually, the algorithm will perform maximal matching, and will return the mapping $\tau$ with the highest match score.

Note that the similarity threshold $\beta$ balance between the mapping precision and recall. Low $\beta$ values will enable more edges in the bipartite graph, which results in higher recall. However, more edges may introduce noise, which in turn will be harmful to the precision. For our use case it is crucial to have a mapping with high precision, hence high $\beta$ values will be used.

\begin{figure}[htb!]
    \centering
    \begin{tikzpicture}[thick, scale=.78,
    ssnode/.style={fill=mygreen,node distance=.1cm,draw,circle, minimum size=.1mm},
    -,shorten >= 2pt,shorten <= 2pt,
    every node/.style = {shape=rectangle, rounded corners, draw, align=center, top color=white, bottom color=blue!20},
    map1/.style={shape=rectangle, rounded corners, draw, align=center, top color=white, bottom color=orange!40, scale=.75},
    map2/.style={shape=rectangle, rounded corners, draw, align=center, top color=white, bottom color=green!40, scale=.75},
    level 1/.style={level distance=.3cm},
    level 2/.style={sibling distance=-19mm},
    level 5/.style={level distance=1cm, sibling distance=2mm},
    level distance=.7cm
    ]
    \node[map1] (b5) at (-4,5) {\scriptsize{(aname, Tova M.)}};
    \node[map1] (b1) at (-4,4) {\scriptsize{(oname, TAU)}};
    \node[map1] (b2) at (-4.2,1.8) {\scriptsize{(ptitle, OASSIS...)}};
    \node[map1] (b3) at (3.5,1.5) {\scriptsize{(cname, SIGMOD)}};
    \node[map1] (b4) at (-3.2,1.2) {\scriptsize{(pyear, 2014)}};
    
    \node[map2] (c5) at (1.5,5) {\scriptsize{(oname, TAU)}};
    \node[map2] (c1) at (1.5,4) {\scriptsize{(aname, Tova M.)}};
    \node[map2] (c4) at (-1.3,1) {\scriptsize{(pyear, 2014)}};

    \begin{scope}[level 1/.style={level distance=1.0cm}, yshift= 7cm, xshift= -1.7cm]

    \Tree 
    [.return
    [.\node[] (a5) {organization}; {the}
    [.\node[] (z) {of};
    [.\node[] (a1) {authors};
    [.{published} {who}
    [.\node[] (a2) {papers}; ]
    [.{after} [.\node[] (a4) {2005}; ] ]
    [.{in}
    [.\node[] (a3) {conferences};
    [.\node[] (a6) {database}; ]
    ]
    ]
    ]
    ]
    ]
    ]
    ]

    \end{scope}

    \draw[->,thick,shorten <=2pt,shorten >=2pt, dashed] (a5) -- (b5);
    \draw[->,thick,shorten <=2pt,shorten >=2pt, dashed] (a5) -- (c5);
    \draw[->,thick,shorten <=2pt,shorten >=2pt, dashed] (a1) -- (b1);
    \draw[->,thick,shorten <=2pt,shorten >=2pt, dashed] (a1) -- (c1);
    \draw[->,thick,shorten <=2pt,shorten >=1, dashed] (a2) -- (b2);
    \draw[->,thick,shorten <=2pt,shorten >=1, dashed] (a3) -- (b3);
    \draw[->,thick,shorten <=2pt,shorten >=1, dashed] (a4) -- (b4);
    \draw[->,thick,shorten <=2pt,shorten >=1, dashed] (a4) -- (c4);
    \end{tikzpicture}
    \caption{Dependency-To-Query Partial Mappings}
    \label{fig:mappings_recall_precision}
\end{figure}
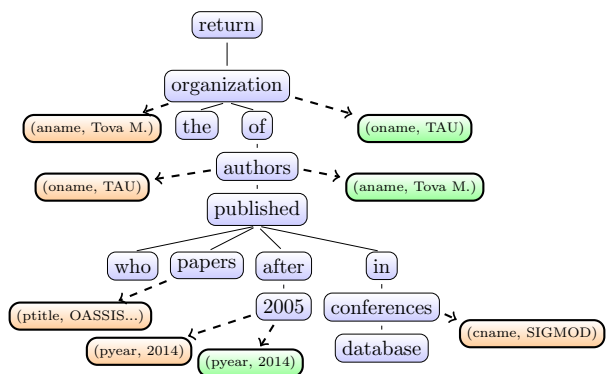

\begin{example}
Recall our running example, and consider the two mapping functions presented in Figure \ref{fig:mappings_recall_precision}.
$\tau_1$ depicted in the orange nodes has high recall, as all of the relevant tree nodes mapped to query parts, however it does not have perfect precision as \emph{organization} node is incorrectly mapped to \emph{aname} and \emph{authors} node is mapped to \emph{oname}. As a result our answer will be:
\\
    \begin{tabular} {l}
        \verb"Tova M. is the organization of TAU who published 'OASSIS...' in SIGMOD in 2014"
    \end{tabular}
\\
This answer makes no sense, and will cause the user to mistrusts the answer and the system.
On the other hand $\tau_2$, depicted in the green nodes, has perfect precision but low recall as \emph{papers} and \emph{conferences} nodes are not mapped to any variable; this will result in:
\\
    \begin{tabular} {l}
        \verb"TAU is the organization of Tova M. who published papers in database conferences in 2014"
    \end{tabular}
\\
Even though the answer does not supply all relevant information, it is a coherent sentence, and clearly a better answer than the previous one.
\end{example}

Since the dependency tree can be artificially made by our system from the NL query, the only recommended component of this NLIDB is a mapping from words of the NL query to the parts of the formal query. Therefore, any NLIDB with such a component could work well with our system (e.g. PRECISE \cite{popescu2003} and ATHENA \cite{SahaFSMMO16}). But even this component can be replaced by Algorithm \ref{algo:mapper} which artificially generates such a mapping. If we do use this algorithm, any NLIDB can be fitted to the system.

\subsubsection{Alternative Provenance Models}
We have used a detailed value-level provenance model for UCQs, which we have leveraged to connect different pieces of the provenance with different parts of the NL question, eventually resulting in a detailed answer. We next briefly discuss alternatives and extensions.

\paragraph{Tuple-level provenance} Using value-level provenance has been essential in our construction of the NL representation of provenance, and consequently in the generation of answers with NL explanations. 

If the system is connected to a system that only allows coarser-grained provenance, then the mapping between words and values needs to be otherwise constructed. Namely, one could use provenance that is only at a tuple-level, as is typically done in provenance models (including the standard semiring model \cite{GKT-PODS07}, why-provenance \cite{why}, lineage \cite{Re2008}, etc.). Then, considering all values in the tuple participating in the provenance, we can reconstruct the mapping to NL query words in alternative means, such as word embeddings and semantic similarity.

\paragraph{Operator-level provenance} An even coarser grain view of the provenance is at the {\em operator level}. For instance, in the context of relational queries one may consider tracking tuples that are the input and output of each operator in the query plan, while not necessarily keeping track of which input contributed to which output. Can such form of provenance be useful in our setting?

One use-case of marrying operator-level provenance with NL queries is in the context of provenance for non-answers. Examining the set of input and output tuples of each operator in the query plan, the work of \cite{ChapmanJ09} defines the notion of a ``picky" operator with respect to a tuple, as one that is responsible for its omission from the output.
This opens up possibilities for explaining non-answers. In a recent preliminary work \cite{deutch2018nlprovenans} we have combined \nalir{}'s mapping of words to query operators, with the work of \cite{ChapmanJ09} to identify the words that map to picky operators. Then, for each requested non-answer, we can highlight this word as ``responsible". 

\begin{figure}[]
\centering
\begin{tikzpicture}
[grow'=up,scale=0.9,level 1/.style={level distance=0.9cm, sibling distance=-20mm}, level 2/.style={sibling distance=-0mm}, level distance=.6cm,
        every node/.style = {shape=rectangle, rounded corners, draw, align=center, top color=white, bottom color=green!20},
        picky/.style={shape=rectangle, rounded corners, draw, align=center, top color=white, bottom color=red!40}]
\Tree 
	[.$\Pi_{oname}$
		[.$\bowtie_{cid}$ 
			[.$\sigma_{dname = databases}$ [.$\bowtie_{cid}$ {\bf{\em conf}} [.$\bowtie_{did}$ {\bf{\em domain}} {\bf{\em domainConf}} ]]]
			[.\node[picky] {$\sigma_{pyear > 2005}$};
				[.$\bowtie_{wid}$ {\bf{\em pub}}
					[.$\bowtie_{aid}$ 
					[.$\bowtie_{oid}$ {\bf{\em author}} {\bf{\em org}} ] {\bf{\em writes}} ] ] ] ] ]
\end{tikzpicture}
\caption{Query Plan with Frontier Picky}\label{fig:plan}
\end{figure}
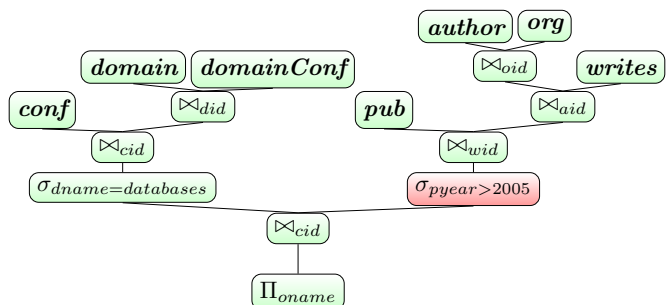

\begin{example}\label{ex:whynot}
Reconsider our running example, but this time assume it is executed on a smaller dirty DB as depicted in Figure \ref{tbl:dblpfaulty}, where the papers ``OASSIS: \ldots'' and ``A sample \ldots'' are erroneously associated with the publication year 2004 instead of 2014. Due to the errors in the database ``TAU'' will not return as an answer to the query, and a user who expects to see ``TAU'' in the results screen will be interested to understand the reason for its absence (and fix the database accordingly).
Consider the query evaluation plan in Figure \ref{fig:plan}, for the query in Figure \ref{fig:nlcq}. The frontier picky operator for ``TAU'' is $\sigma_{pyear} > 2005$, thus the system depicted in \cite{deutch2018nlprovenans} will highlight the relevant part in the NL query, and return
\\
    \begin{tabular} {l}
        \verb"return the organization of authors who published papers in database conferences"\\ \textbf{\underline{after 2005.}}\\
    \end{tabular}
\\
Indicating ``TAU'' is a non-answer because the authors associated with it did not published papers after 2005.
\end{example}

Similarly to the approach discussed here, the system utilizes the mappings from words to operators, constructed by the NLIDB, and highlights the relevant term which filtered the queried tuple. A challenge arises when the filtering operator has no direct word or phrase in the NL query mapped to it. 

\begin{example}
Continuing Example \ref{ex:whynot}, if the query was about an organization whose authors did not publish in any database conference after 2005, the filtering operator would have been the join between the $author$ and $writes$ tables. Since there is no direct mapping between a word in the NL query and the join operator, it is unclear which word/phrase to highlight. 
\end{example}

\begin{figure}[!htb]
    \centering \scriptsize
    \begin{minipage}{.5\linewidth}
        \centering
        \caption*{Rel. $\text{\textit{org}}$}\label{tbl:organizationfaulty}
        \begin{tabular}{| c | c | c | c | c | c |}
            \hline oid & oname \\
            \hline 1 & UPENN \\
            \hline 2 & TAU \\
            \hline
        \end{tabular}
        
    \end{minipage}%
    \begin{minipage}{.5\linewidth}
        \centering
        \caption*{Rel. $\text{\textit{author}}$}\label{tbl:authorfaulty}
        \begin{tabular}{| c | c | c | c | c | c |}
            \hline aid & aname & oid \\
            \hline 3 & Susan D. & 1 \\
            \hline 4 & Tova M. & 2 \\
            \hline 5 & Slava N. & 2 \\
            \hline
        \end{tabular}
        
    \end{minipage}

    \begin{minipage}{0.7\linewidth}
        \centering
        \caption*{Rel. $\text{\textit{pub}}$}\label{tbl:publicationfaulty}
        \begin{tabular}{| c | c | c | c | c | c |}
            \hline wid & cid & ptitle & pyear\\
            \hline 6 & 10 & ``OASSIS..." & \cellcolor{red!40}2004 \\
            \hline 7 & 10 & ``A sample..." & \cellcolor{red!40}2004 \\
            \hline
        \end{tabular}
        
    \end{minipage}%
    \begin{minipage}{.3\linewidth}
        \centering
        \caption*{Rel. $\text{\textit{writes}}$}\label{tbl:writesfaulty}
        \begin{tabular}{| c | c | c | c | c | c |}
            \hline aid & wid\\
            \hline 4 & 6 \\
            \hline 3 & 6 \\
            \hline 5 & 6 \\
            \hline 4 & 7 \\
            \hline
        \end{tabular}
        
    \end{minipage}

    \begin{minipage}{.33\linewidth}
        \centering
        \caption*{Rel. $\text{\textit{conf}}$}\label{tbl:conferencefaulty}
        \begin{tabular}{| c | c | c | c | c | c |}
            \hline cid & cname\\
            \hline 10 & SIGMOD \\
            \hline
        \end{tabular}
        
    \end{minipage}
    \begin{minipage}{.33\linewidth}
        \centering
        \caption*{Rel. $\text{\textit{domainConf}}$}\label{tbl:domainconferencefaulty}
        \begin{tabular}{| c | c | c | c | c | c |}
            \hline cid & did\\
            \hline 10 & 18 \\
            \hline
        \end{tabular}
        
    \end{minipage}%
    \begin{minipage}{.33\linewidth}
        \centering
        \caption*{Rel. $\text{\textit{domain}}$}\label{tbl:domainfaulty}
        \begin{tabular}{| c | c | c | c | c | c |}
            \hline did & dname\\
            \hline 18 & Databases \\
            \hline
        \end{tabular}
        
    \end{minipage}

    \caption{Faulty DB Instance}\label{tbl:dblpfaulty}
\end{figure}

\paragraph{Provenance Beyond UCQs}
A limitation of our work is that it is limited to the SPJU fragment of SQL (UCQs), while NLIDBs have considerable success in handling questions that compile to far more expressive formalisms. \nalir{} in particular also supports nesting and aggregation, both lacking support in our solution.

Provenance models for such formalisms do exist, from \cite{Amsterdamer2011} and \cite{Olteanu2012} for aggregate queries, \cite{KwasnikowskaB08} for nested queries, \cite{LeeKLG17} for queries with negation to \cite{glavic2014big} for full SQL, just as few examples. By and large, these solutions are intended as internal representations. Presenting them as explanations is an important task that lack satisfactory solutions. For instance, the model of \cite{Amsterdamer2011} for aggregate results includes in a sense a record of all tuples participating in the aggregate computation, which may be far too many to show to the user. The work of \cite{Olteanu2012} discusses a factorized circuit-based form, yet it is also too complicated to allow its presentation to a user. We view devising an effective way of showing such provenance instances to users -- e.g. summarizing the contribution of individual tuples in aggregate queries -- as an important goal for future work.     
\section{Experiments}
 \label{sec:experiments}

We have performed an experimental study to assess \systemName\ through two prisms:
(1) the quality of answers produced by the system, and
(2) the efficiency of the algorithms in terms of execution time.

\subsection{User Study}

\begin{table}[]
    \centering \scriptsize
    \caption{NL queries}\label{tbl:queries}
    \begin{tabular}{| C{0.7cm} | p{7cm} | }
        \hline {\bf Num.} & {\bf Queries} \\
        \hline
        1 & \verb"Return the homepage of SIGMOD" \\
        \hline 2 & \verb"Return the papers whose title contains 'OASSIS'" \\
        \hline 3 & \verb"Return the papers which were published in"
        \verb"conferences in database area" \\
        \hline 4 & \verb"Return the authors who published papers in"
        \verb"SIGMOD after 2005" \\
        \hline 5 & \verb"Return the authors who published papers in"
        \verb"SIGMOD before 2015 and after 2005" \\
        \hline 6 & \verb"Return the authors who published papers in"
        \verb"database conferences" \\
        \hline 7 & \verb"Return the organization of authors who published"
        \verb"papers in database conferences after 2005" \\
        \hline 8 & \verb"Return the authors from TAU who published"
        \verb"papers in VLDB" \\
        \hline 9 & \verb"Return the area of conferences" \\
        \hline 10 & \verb"Return the authors who published papers in"
        \verb"database conferences after 2005" \\
        \hline 11 & \verb"Return the conferences that presented papers"
        \verb"published in 2005 by authors from organization" \\
        \hline 12 & \verb"Return the years of paper published"
        \verb"by authors from IBM" \\
        \hline 13 & \verb"Return the authors who published papers in"
        \verb"VLDB or SIGMOD after 2005" \\
        \hline 14 & \verb"Return the authors from TAU or HUJI who published"
        \verb"papers in VLDB or SIGMOD" \\
        \hline 15 & \verb"Return the papers published by authors from"
        \verb"TAU or HUJI" \\        
        \hline
    \end{tabular}
    
\end{table}

\begin{table*}[ht!]
    \centering\scriptsize
    \caption{Sample use-cases and results}\label{tbl:usecases}
    \begin{tabular}{| p{4.5cm} | p{4cm} | p{6cm} |}
        \hline \textbf{Query} & \textbf{Single Assignment} & \textbf{Multiple Assignments - Summarized}\\
        \hline Return authors from TAU & Tova M. from TAU & \\
        \hline Return the homepage of SIGMOD & http://www.sigmod2011.org/ is the homepage of SIGMOD & \\
        \hline Return the domain of VLDB & Databases is the domain of VLDB & \\
        \hline Return the domain of conferences & Databases is the domain of VLDB & {Databases is the domain of 260 conferences} \\
        \hline Return the year of VLDB paper & 2007 is the year of VLDB ``Graph Partitioning..." paper & {2007 is the year of VLDB 152 papers} \\
        \hline Return authors who published in papers in a journal & Tova M. published ``Putting lipstick on Pig..." in CORR & {Tova M. published 60 papers in 17 journals}\\        
        \hline Return the authors who published papers in SIGMOD before 2015 and after 2005 & Tova M. published ``Auto-completion..." in SIGMOD in 2012 & {Tova M. published 10 papers in SIGMOD in 2006--2014}\\
        \hline Return the authors from TAU who published papers in VLDB & Tova M. from TAU published ``XML Repository..." in VLDB & {Tova M. from TAU published 11 papers in VLDB}\\
        \hline Return the authors who published papers in database conferences & Tova M. ``published Auto-completion..." in SIGMOD & {Tova M. published 96 papers in 18 conferences}\\
        \hline Return the organization of authors who published papers in database conferences after 2005 & TAU is the organization of Tova M. who published `OASSIS...' in SIGMOD in 2014 & {TAU is the organization of 43 authors who published 170 papers in 31 conferences in 2006 - 2015}\\
        \hline Return the authors who published papers in VLDB or SIGMOD after 2005 & Tova M. published ``Auto-completion..." in SIGMOD in 2012 & {Tova M. published 12 papers in VLDB or SIGMOD in 2006--2014}\\
        \hline
    \end{tabular}
    
\end{table*}

We have examined the usefulness of the system through a user study,
involving 22 non-expert users. 
The user study was conducted in two phases, first we asked 15 users to evaluate the solution for SPJ queries, where in the second phase 7 different users were requested to evaluate the solution for union queries. 
For the SPJ evaluation we presented to each user 6 NL queries, namely No. 1--4, 6, and 7 from Table \ref{tbl:queries} 
(chosen as a representative sample), where for the union evaluation users were presented with queries 13--15. We have also allowed
each user to freely formulate an NL query of her choice, related to
the MAS database \cite{mas}. 2 users have not provided a query at all, and for 5 users the query either did not parse well or involved
aggregation (which is not supported), leading to a total of 119
successfully performed tasks. For each of the NL queries, users were
shown the NL provenance computed by \systemName\ for cases of single
derivations, factorized and summarized answers for multiple
derivations (where applicable). Multiple derivations were relevant
in 71 of the 119 cases. Examples of the results are shown
in Table \ref{tbl:usecases}.

We have asked users three questions about each case, asking them to
rank the results on a 1--5 scale where 1 is the lowest score: (1) is
the answer relevant to the NL query? (2) is the answer
understandable? and (3) is the answer detailed enough, {\textit i.e.} supply
all relevant information? (asked only for answers including multiple
assignments).

The results of our user study are summarized in Figure
\ref{tbl:userStrudy}. In all cases, the user scores were in the
range 3--5, with the summarized explanation receiving the highest
scores on all accounts. Note in particular the difference in
understandability score, where summarized sentences ranked as
significantly more understandable than their factorized
counterparts. Somewhat surprisingly, summarized sentences were even
deemed by users as being more detailed than factorized ones
(although technically they are of course less detailed), which may
be explained by their better clarity (users who ranked a result
lower on understandability have also tended to ranked it low w.r.t.
level of detail).

\begin{figure}[]
    \centering \scriptsize
    \begin{tabular}{| p{1.8cm} | c | c | c | c | c | c | c | c | c |}
        \hline
        \rowcolor[HTML]{C0C0C0} 
        & \multicolumn{4}{|c|}{SPJ Queries} & \multicolumn{4}{|c|}{Union Queries}\\
        \hline
        \hline 
        \rowcolor[HTML]{D7D7D7}
        {\bf Category} & {\bf 3} & {\bf 4} & {\bf 5} & {\bf Avg.} & {\bf 3} & {\bf 4} & {\bf 5} & {\bf Avg.} \\
        \hline
        \hline \rowcolor[HTML]{F0F0F0} \multicolumn{9}{|l|}{\textbf{Single}}\\
        \hline Relevant & 4 & 10 & 84 & \textbf{4.82} & 0 & 5 & 16 & \textbf{4.76} \\
        \hline Understandable & 7 & 25 & 66 & \textbf{4.60} & 0 & 4 & 17 & \textbf{4.81} \\
        \hline
        \hline \rowcolor[HTML]{F0F0F0} \multicolumn{9}{|l|}{\textbf{Multiple}}\\
        \hline Relevant & 0 & 7 & 43 & \textbf{4.86} & 0 & 6 & 15 & \textbf{4.71} \\
        \hline Understandable & 4 & 13 & 33 & \textbf{4.58} & 1 & 7 & 13 & \textbf{4.57} \\
        \hline Detailed & 3 & 7 & 40 & \textbf{4.74} & 0 & 7 & 14 & \textbf{4.67} \\
        \hline
        \hline \rowcolor[HTML]{F0F0F0} \multicolumn{9}{|l|}{\textbf{Summarized}}\\
        \hline Relevant & 2 & 2 & 46 & \textbf{4.88} & 0 & 4 & 17 & \textbf{4.81} \\
        \hline Understandable & 3 & 3 & 44 & \textbf{4.82} & 0 & 3 & 18 & \textbf{4.86} \\
        \hline Detailed & 2 & 5 & 43 & \textbf{4.82} & 0 & 6 & 15 & \textbf{4.71} \\
        \hline
    \end{tabular}
    \caption{Users ranking}\label{tbl:userStrudy}
\end{figure}

\subsection{Scalability}\label{subsec:scale}

Another facet of our experimental study includes runtime experiments
to examine the scalability of our algorithms. Here again we have
used the MAS database whose total size is 4.7 GB, and queries No. 1--15 from Table
\ref{tbl:queries}, running the algorithm to generate NL provenance
for each individual answer. The experiments were performed on a i7
processor and 32GB RAM with Windows 8. As expected, when the
provenance includes a single assignment per answer, the runtime is negligible (this is the case for queries No. 
1--3). We thus show the results only for queries No. 4--15.

\begin{table}[!htb]
    \centering \scriptsize
    \caption{Computation time (sec.), for the MAS database}\label{tbl:masTimes}
    \begin{tabular}{| c | c | c | c | c |c | c | c |}
        \hline {\bf Query} & \begin{tabular}{@{}c@{}} {\bf Query Eval.} \\ {\bf Time} \end{tabular} & \begin{tabular}{@{}c@{}} {\bf Fact.} \\ {\bf Time} \end{tabular} & \begin{tabular}{@{}c@{}} {\bf Sentence} \\ {\bf Gen. Time}
        \end{tabular} & \begin{tabular}{@{}c@{}} {\bf \systemName} \\ {\bf Time} \end{tabular}\\
        \hline 4 & 0.9 & 0.038 & 0.096 & 0.134 \\
        \hline 5 & 0.6 & 0.03 & 0.14 & 0.17 \\
        \hline 6 & 33 & 0.62 & 2.08 & 2.7 \\
        \hline 7 & 20.5 & 1.1 & 3.1 & 4.2 \\
        \hline 8 & 2.4 & 0.001 & 0.001 & 0.002 \\
        \hline 9 & 0.01 & 0.011 & 0.001 & 0.012 \\
        \hline 10 & 21.3 & 0.53 & 2.23 & 2.76 \\
        \hline 11 & 53.7 & 3.18 & 6.46 & 9.64 \\
        \hline 12 & 18.8 & 3.22 & 1.73 & 4.95 \\
        \hline 13 & 1.4 &  0.07 & 0.33 & 0.4 \\
        \hline 14 & 14.4 & 0.001 & 0.004 & 0.005 \\
        \hline 15 & 5.5 & 0.1 & 0.41 & 0.51 \\
        \hline
    \end{tabular}
    
\end{table}

Table \ref{tbl:masTimes} includes, for each query, the runtime
required by our algorithms to transform provenance to NL in
factorized or summarized form, for all query results (as explained
in Section \ref{sec:multiple}, we can compute the factorizations
independently for each query result). We show a breakdown of the
execution time of our solution: factorization time, sentence
generation time, and total time incurred by \systemName\ (we note
that the time to compute summarizations given a factorization was
negligible). For indication on the complexity level of the queries,
we also report the time incurred by standard (provenance-oblivious)
query evaluation, using the mySQL engine. 
We note that our algorithms perform quite well for all queries 
(overall \systemName\ execution has $15\%$ overhead), 
even for fairly complex ones such as queries 7, 11, and 12.

\begin{figure*}[!htb]
\captionsetup{justification=centering}
    \hspace*{-0.3cm}
    \centering
    \begin{subfigure}[b]{0.33\textwidth}
        \includegraphics[width=\textwidth]{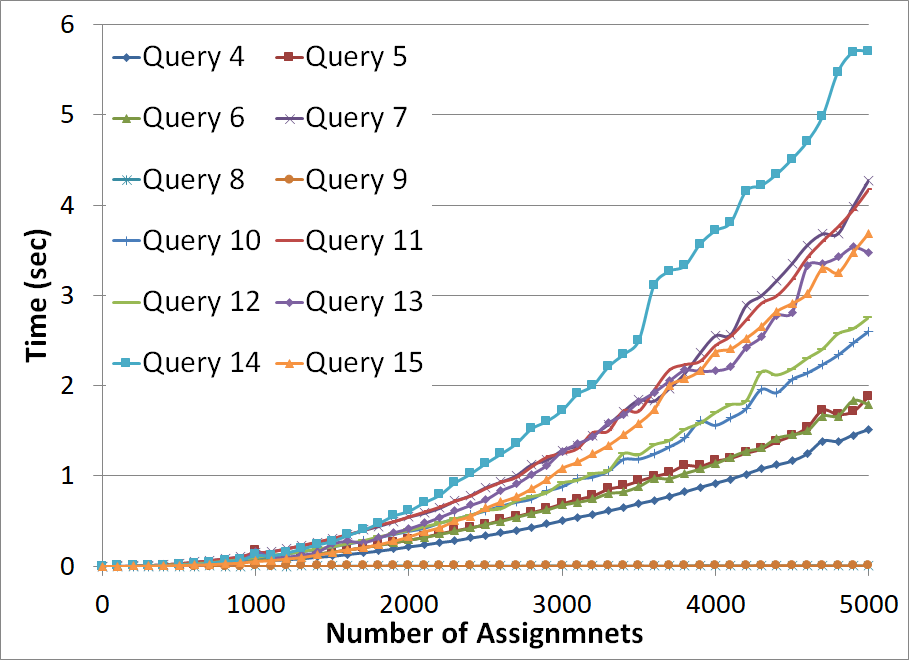}
        \caption{Computation time as a function of the number of assignments}
        \label{graph:uniqueRuntime}
    \end{subfigure}\hfill%
    \begin{subfigure}[b]{0.33\textwidth}
        \includegraphics[width=\textwidth]{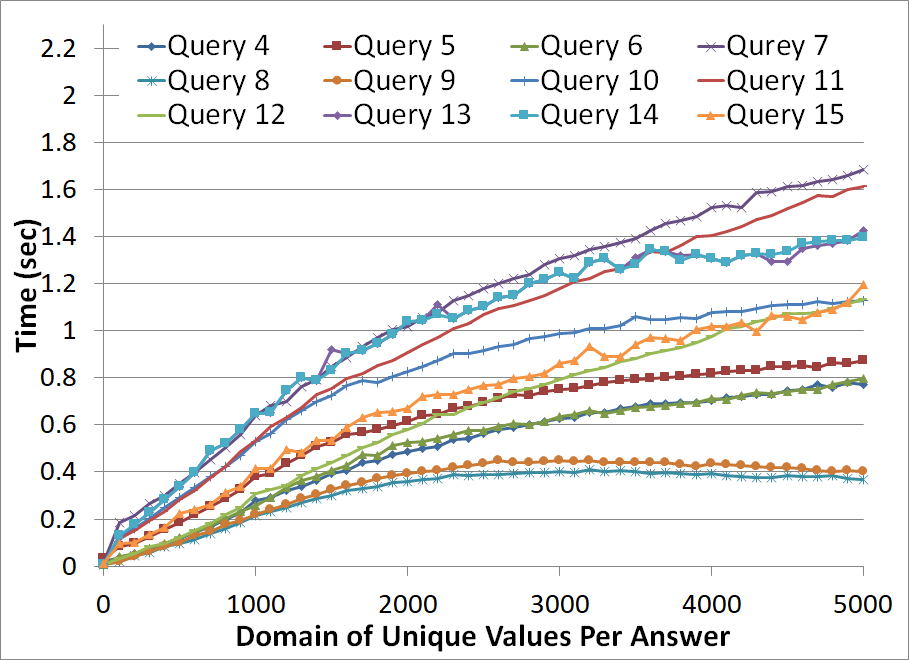}
        \caption{Computation time as a function of the number of unique values}
        \label{graph:sharedRuntime}
    \end{subfigure}\hfill%
    \begin{subfigure}[b]{0.33\textwidth}
        \includegraphics[width=\textwidth]{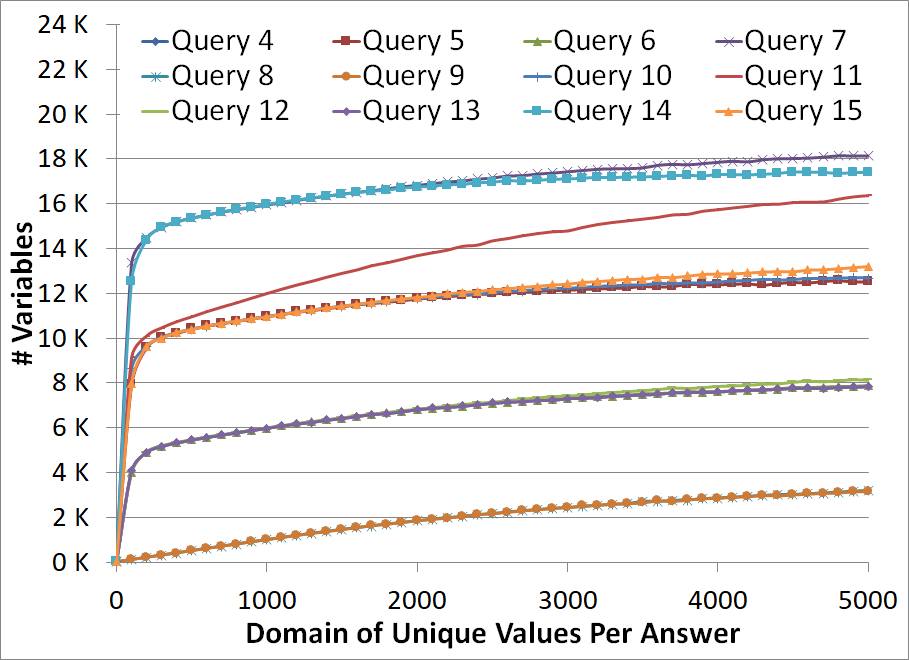}
        \caption{Factorization size as a function of the number of unique values}
        \label{graph:factSize}
    \end{subfigure}
    \caption{Results for synthetic data}
\end{figure*}

Figure \ref{graph:uniqueRuntime} (see next page) presents the execution time of NL
provenance computation for an increasing number of assignments {\em
per answer} (up to 5000, note that the maximal number in the real data experiments was 4208). The provenance used for this set of experiments
was such that the only shared value in all assignments was the
result value, so the factorization phase is negligible in terms of
execution time, taking only about one tenth of the total runtime in
the multiple assignments case. Most computation time here is
incurred by the answer tree structuring. We observe that the
computation time increased moderately as a function of the number of
assignments (and is negligible for the case of a single assignment).
The execution time for 5K assignments with unique values was 1.5, 2,
1.9, 4.9, 0.006, 0.003, 2.6, 5.3, 3.7, 3.5, 5.7, and 3.7 seconds resp. for
queries 4--15. Summarization time was negligible, less than
0.1 seconds in all cases.

\begin{figure}[!htb]
    \centering
    \begin{subfigure}[b]{0.33\textwidth}
        \includegraphics[width=\textwidth]{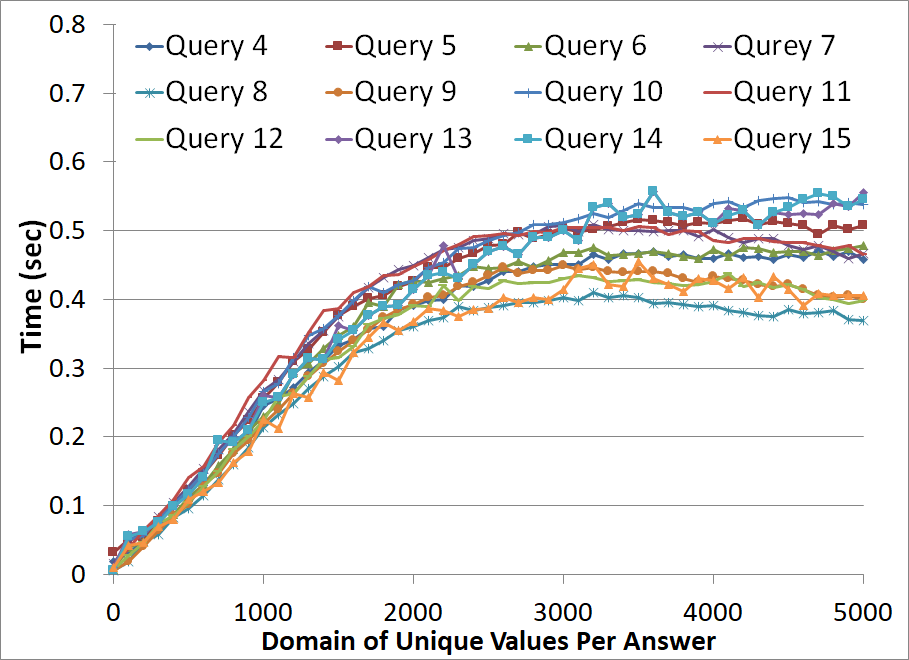}
        \caption{Factorization time}
        \label{graph:factTime}
    \end{subfigure}
    \begin{subfigure}[b]{0.33\textwidth}
        \includegraphics[width=\textwidth]{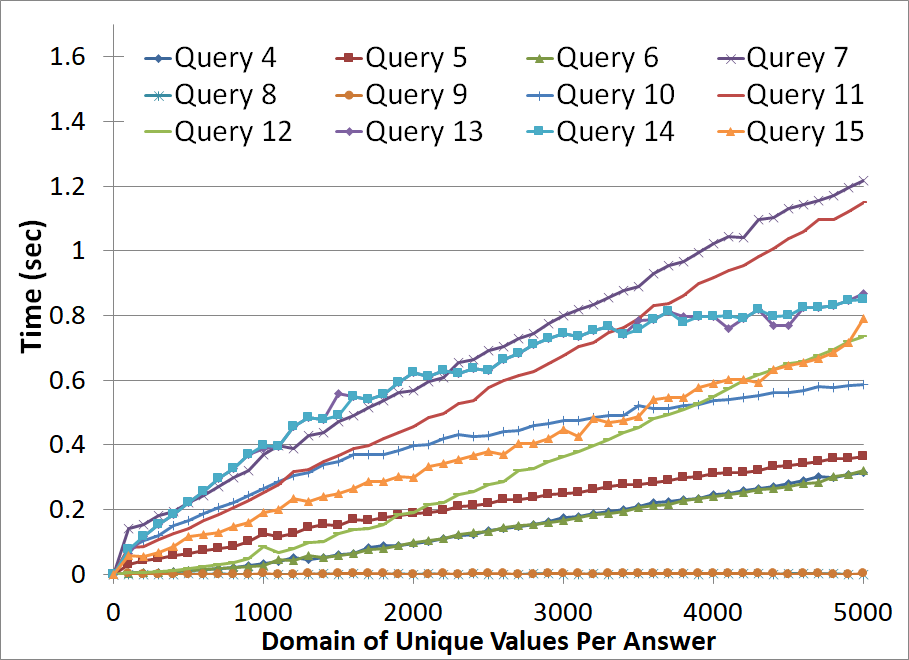}
        \caption{Sentence gen. time}
        \label{graph:sentenceTime}
    \end{subfigure}
    \caption{Breakdown for synthetic experiments}
    \label{fig:breakdown}
\end{figure}

For the second set of experiments, we have fixed the number of
assignments per answer at the maximum 5K and changed only the domain
of unique values from which provenance expressions were generated.
The domain size {\em per answer, per query variable} varies from 0 to 5000 (this cannot exceed the number of assignments).
Note that the running time increases as a function of the number of unique values:
when there are more unique values, there are more candidates for
factorization (so the number of steps of the factorization algorithm
increases), each factorization step is in general less effective (as
there are more unique values for a fixed size of provenance, {\textit i.e.}
the degree of value sharing across assignments decreases), and
consequently the resulting factorized expression is larger, leading
to a larger overhead for sentence generation. Indeed, as our
breakdown analysis (Figure \ref{fig:breakdown}) shows, the increase
in running time occurs both in the factorization and in the sentence
generation time. Finally, Figure \ref{graph:factSize} shows the
expected increase in the factorized expression size w.r.t the
number of unique values.

\begin{figure}[!htb]
    \centering
    \begin{subfigure}[b]{0.3\textwidth}
        \centering
        \includegraphics[width=\textwidth]{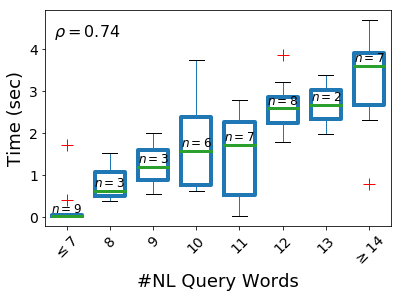}
        \caption{}
        \label{graph:aggWords}
    \end{subfigure}
    \begin{subfigure}[b]{0.3\textwidth}
        \centering
        \includegraphics[width=\textwidth]{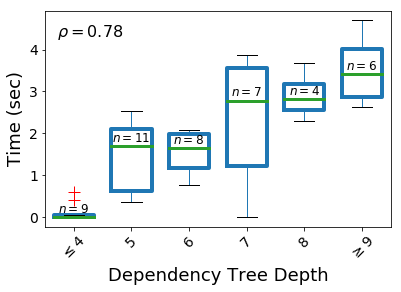}
        \caption{}
        \label{graph:aggDepth}
    \end{subfigure}
    \begin{subfigure}[b]{0.3\textwidth}
        \centering
        \includegraphics[width=\textwidth]{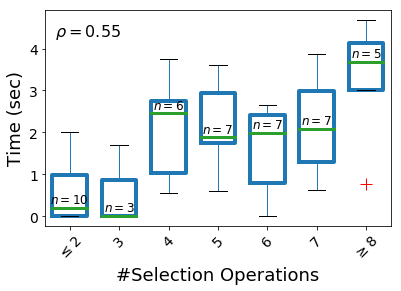}
        \caption{}
        \label{graph:aggSelections}
    \end{subfigure}
    
    \begin{subfigure}[b]{0.3\textwidth}
        \centering
        \includegraphics[width=\textwidth]{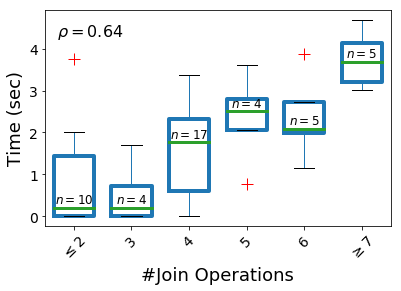}
        \caption{}
        \label{graph:aggJoins}
    \end{subfigure}
    \begin{subfigure}[b]{0.3\textwidth}
        \centering
        \includegraphics[width=\textwidth]{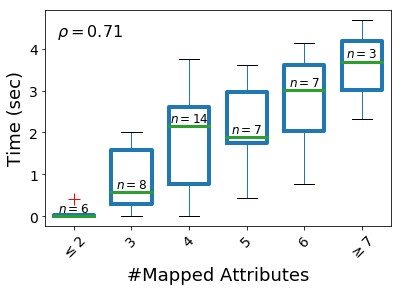}
        \caption{}
        \label{graph:aggAttributes}
    \end{subfigure}
    \caption{Computation time as function of \textbf{(a)} NL query length \textbf{(b)} depth of the NL query dependency tree \textbf{(c)} number of query selection operations \textbf{(d)} number of query join operations \textbf{(e)} number of provenance attributes}
    \label{fig:agg}
\end{figure}

For the third scalability experiment we evaluated the computation time for different classes of queries. In this experiment we have used a larger set of queries, consisting of 45 different NL queries (available in \cite{git}) which vary in their size, structure, and complexity.
For each query we have fixed both the number of assignments per answer and the domain of unique values at 5K.
\systemName{} computation time varied between $0.005$ to $4.69$ seconds, where the mean and median computation times were $1.73$ and $1.8$ seconds respectively.
Figure \ref{fig:agg} depict aggregation of the computation times with respect to different query aspects. 
Figures \ref{graph:aggWords} and \ref{graph:aggDepth} explore the influence of the original NL query sentence structure, recall that both Algorithms \ref{algo:ansTree} and \ref{algo:factorizedTree} utilize the query tree in order to generate the answer sentence, hence the computation time increase as function of the query sentence length~\textbackslash~tree depth, with Pearson correlation of 0.74 and 0.78 respectively.
The impact of the formal query complexity on \systemName{} running time is presented in Figures \ref{graph:aggSelections} and \ref{graph:aggJoins}, notice that the query complexity influence the query evaluation time, but does not have direct impact on the explanation generation, hence a lower correlation was measured (0.55 and 0.64 for number of selection and join operations respectively). 
Finally, Figure \ref{graph:aggAttributes} depict the impact of provenance size on the computation time, the number of provenance attributes is crucial for the factorization step, hence it is influential on Algorithm \ref{algo:factorizedTree} running time as exhibited by the 0.71 correlation.
\section{Related Work} \label{sec:relatedWork}
In this section we review and compare our work to existing approaches in the context of database theory and database interfaces.

\paragraph{Provenance}
The tracking, storage and presentation of provenance have been the
subject of extensive research in the context of database queries,
scientific workflows, and others (see {\textit e.g.}
\cite{why,Herschel,GKT-PODS07,CheneyProvenance,Greenicdt09,DavidsonF08,DavidsonBELMBAF07,glavic2013using,KenigGS13}) 
while the field of provenance applications has also been broadly
studied ({\textit e.g.} \cite{vldb15deutch,MeliouSS12,RoyS14}). A longstanding
challenge in this context is the complexity of provenance
expressions, leading to difficulties in presenting them in a
user-comprehensible manner. Approaches in this respect include
showing the provenance in a graph form
\cite{karma,Missier2010,Taverna,chimera,DavidsonF08,CohnH09,AilamakiIL98}, allowing user control over the level of granularity (``zooming" in
and out \cite{Cohen2008}), or otherwise presenting different ways of provenance visualization \cite{Herschel}.
Other works have studied
allowing users to query the provenance ({\textit e.g.}
\cite{proql,tao2012querying}) or to a-priori request that 
only parts of the provenance are tracked (see for example
\cite{vldb15deutch,glavic2014big,GlavicICDE09}). Importantly
provenance factorization and summarization have been studied ({\textit e.g.},
\cite{Chapman2008,factorize,Olteanu2012,Re2008}) as means for
compact representation of the provenance. Usually, the solutions
proposed in these works aim at reducing the size of the provenance
but naturally do not account for its presentation in NL; we have
highlighted the different considerations in context of
factorization/summarization in our setting. We note that {\em
value-level provenance} was studied in \cite{valueProv,DBNotes} to
achieve a fine-grained understanding of the data lineage, but again
do not translate the provenance to NL.

\paragraph{Detailed Answers to Keyword Queries}
There is an extensive line of work on answering keyword queries which focuses on providing not just the query answer (tuples that contain the queried value), but also comprehensive details about it. Works such as \cite{2002DBXplorer,BhalotiaHNCS02,HristidisP02} focus on answering keyword queries over a relational database, by outputting tuples that are related to one or more of the queried keywords. 
In particular, \cite{KoutrikaSI06,SimitsisKI08} studies the subject of pre\'{c}is queries over relational database. These queries are logical combinations of keywords. The query along with constraints on the schema is inputted to the system, and the answer returned should include the most relevant tuples to the keyword(s), according to the constraints, as relations that form a logical subset of the original database (i.e., contain not only items directly related to the given query terms but also items implicitly related to them). 
Still in the field of answering keyword queries, \cite{Fakas08,FakasCM16,Fakas2014} deal with snippets of a database object, which is an entity that has its identity in the result tuple. In this scenario, the system provides a snippet which is a summary of the relevant information related to these objects. This information is taken from tuples that relate to the queried object's tuple and is prioritized in different manners (e.g., diversity and proportionality). 
All of these works, similarly to ours, provide the query answer along with further details about it, these details stem from tuples that relate in some defined manner to the answer.
While there is a commonality between these works and ours, our work supports complex CQs formulated in NL as opposed to keyword queries. 
Answers to keyword queries are not always explicitly specified in the query, e.g., we can ask about and author and get the name of her organization, or get tuples that contain this author from different tables. For CQs, users explicitly specify the form of answer they would like, from which relation they would like it, and what conditions it has to satisfy. 
Additionally, our work defines the related tuples by their membership in the provenance of the result, i.e, the query structure (formulated by the user) is a major factor in determining which tuples will be included in the explanation. Furthermore, our system generates the NL explanation based on the NL query given by the user, and not a textual template.

\paragraph{Summarization of Database Content}
There have been previous works that proposed a summarized presentation of the query results. In the context of keyword queries, the approach of \cite{FakasCM16} gives short summaries on information regarding data object by limiting the number of related tuples (according to the schema), showing only the highest ranked. The summary is represented as a tree where the root is a tuple containing the keywords and the neighboring tuples are the related ones. 
In the context of top aggregate queries, \cite{WenZRY18} presents an approach that summarizes the results using clustering of the top ranked results, by formulating the clustering problem as an optimization problem. The framework of this work is interactive and allows users to choose the number of clusters and other parameters. After observing the results, users can update the parameters and get different results. The summarization aims to serve as an overview of all query results through a summarized relation. Another related approach is \cite{JoglekarGP15} which devised the smart drill-down operator. the operator allows users to obtain interesting summarizations of the tuples in the relation. The summary is essentially the top-k clusters, according to a goal function, of tuples with don't-care values. 
Similarly to these approaches, we focus on tuples with the same values or similar values and 
The SaintEtiQ system \cite{Saint-PaulRM07} summarizes entire database relations using background knowledge with a vocabulary to translate raw tuple values. Summarization is done using a clustering approach. Like our system, there is a notion of getting a more detailed and precise summary containing more information, and a less detailed one which is more compact. 
Provenance summarization has also been proposed by \cite{ainy}, yet it offers an approximated summary of the provenance based on distance, semantic constraints and size, with a possible loss of information. 
Our summarization technique compacts all the tuples in the provenance, through functions like SUM and RANGE, as opposed to showing/summarizing a few representative tuples. We base this summarization on the factorization of the provenance done as an initial step. Furthermore, we focus here on UCQs and do not cover aggregate queries. Finally, we do not rely on background knowledge of a vocabulary, or other external constraints to summarize, but rather use the provenance factorization. Additionally, we translate the summarization into NL which is geared towards non-expert users.

\paragraph{NL Interfaces}
Multiple lines of work ({\textit e.g.} \cite{Kupper1993,Amsterdamer:2015,nalir,SongSSBZBMDDMH15,song:15a,2002DBXplorer,popescu2003}) have proposed NL interfaces for the formulation of database queries, and additional works \cite{franconi} have focused on presenting the {\em answers} in NL, typically basing their translation on the schema of the output relation. Among these, works such as \cite{Amsterdamer:2015,nalir} also harness the dependency tree 
in order to make the translation form NL to SQL by employing {\em mappings} from the NL query to formal terms. The work of \cite{koutrika2010} has focused on the complementary problem of translating SQL {\em queries} (rather than their answers or provenance) to NL. Another work has devised an interactive chatbot interface to drill down and zoom-in on a specific part of the database which the user is interested in \cite{SellamK16}. This work helps guide the user with NL but does not show the query answers and their explanations in natural language. 
In the context of answering keyword queries, \cite{SimitsisK06} shows an approach that presents the results of pre\'{c}is queries as a narrative text so that the output is more user friendly. To do so, there is a need for predefined textual templates to embed the relevant tuples in. The templates are predefined by a designer or the administrator of the database.
Synthesizing text directly from databases has also been explored in \cite{SimitsisAKI08} which extended \cite{SimitsisK06}. This work revolves around the generation of textual representation for database subsets. The text is generated based on templates rather than on user formulated queries in NL. Moreover, the explanation is composed of tuples from related database tables, as opposed to tuples from the provenance which provide a targeted explanation tailored to the specific details provided by a user in an NL query. 
To our knowledge, no previous work has focused on formulating the {\em provenance} of output tuples in NL. This requires fundamentally different techniques ({\textit e.g.} that of factorization and summarization, building the sentence based on the input question structure, etc.)
and leads to answers of much greater detail.

\section{Conclusion}
\label{sec:conc}

We have studied in this paper, for the first time to our knowledge,
provenance for NL queries. We have devised 
a novel model of ``word-to-provenance" mapping, thereby leveraging the structure of the original NL question for the generation of a new NL sentence that captures the answers along with their provenance-based explanations. Since there may be many explanations, even for a single answer, we have developed factorization and summarization techniques that are geared towards sentence generation, showing that they result in new criteria for preferring one factorized/summarized form over another. We have implemented the approach and demonstrated its effectiveness through use cases and experiments.

Our work presented a simple yet effective approach of generating NL explanations based on the user NL query. We have demonstrated that by applying basic transformations on the original question we are able to get understandable and relevant NL explanations. 
Usage of more advanced \emph{Natural Language Generation} techniques can farther improve the explanations quality; this is an interesting direction for future work.

Our implementation is based on a particular NL interface to Databases and on a particular provenance model for UCQs, but we have also discussed at some depth the extension of our solution beyond these settings. This discussion provides indication of the generic nature of the approach, but further research is required to fully realize its potential in these other settings. In particular, we believe that the need to handle more complex queries with nesting, aggregation etc. may lead to new and exciting research avenues.

\paragraph{Acknowledgements} This research was partially supported
by the Israeli Science Foundation (ISF) and by the European Research Council (ERC) under the European Unions Horizon 2020 research and innovation programme (Grant agreement No. 804302).

\bibliographystyle{abbrv}
\bibliography{bibShort}

\end{document}